\documentclass[aps,prx,twocolumn,showpacs,superscriptaddress,floatfix,amsmath,amssymb,longbibliography]{revtex4-2}

\usepackage[titletoc,toc,title]{appendix}
\usepackage{graphicx}
\usepackage{bm}
\usepackage{xcolor}
\usepackage{booktabs}
\usepackage{hyperref}
\usepackage{diagbox}
\usepackage{braket}
\usepackage{bbm} % for bold 1 in appendix

\usepackage[color=green!60]{todonotes}
\usepackage{hyperref}
\hypersetup{
    colorlinks=true,
    linkcolor=blue,
    filecolor=magenta,      
    urlcolor=blue,
    citecolor=blue,
}
\renewcommand{\vec}[1]{\bm{#1}}
\renewcommand{\S}{\mathcal{S}}
\newcommand{\U}{\mathcal{U}}
\newcommand{\R}{\mathcal{R}}
\renewcommand{\H}{\mathcal{H}}

\begin{document}

\title{Multi-modal spectroscopy of order parameter distributions}

\author{Stephen Carr}
\affiliation{Department of Physics, Brown University, Providence, Rhode Island 02912-1843, USA}
\affiliation{Brown Theoretical Physics Center, Brown University, Providence, Rhode Island 02912-1843, USA}
\author{Ilija K.\ Nikolov}
\affiliation{Department of Physics, Brown University, Providence, Rhode Island 02912-1843, USA}
\author{Rong Cong}
\affiliation{Department of Physics, Brown University, Providence, Rhode Island 02912-1843, USA}
\author{Adrian {Del~Maestro}}
\affiliation{Department of Physics and Astronomy, University of Tennessee, Knoxville, TN 37996, USA}
\affiliation{Min H. Kao Department of Electrical Engineering and Computer Science, University of Tennessee, Knoxville, TN 37996, USA}
\author{Chandrasekhar Ramanathan}
\affiliation{Department of Physics and Astronomy, Dartmouth College, Hanover, NH 03755, USA}
\author{V. F. Mitrovi\ifmmode \acute{c}\else \'{c}\fi{}}
\affiliation{Department of Physics, Brown University, Providence, Rhode Island 02912-1843, USA}

% We explicitly provide protocols which allow one to distinguish magnetic versus charge noise.
\date{\today}
\begin{abstract}
    We present a multi-modal spectroscopic paradigm that enables independent measurement of charge and spin degrees of freedom (DOF) in strongly correlated materials.
    This spin-based technique probes symmetry-specific Hamiltonian parameters by analyzing how the time delay between applied pulses ($\tau$)  affects the response.
    We demonstrate ways in which charge DOF that couple through the quadrupolar interaction (inversion symmetric) can be independently measured even in the presence of large magnetic noise (inversion asymmetric).
    The method quantifies both the strength of the interactions and its distribution (noise).
    We provide protocols to directly and independently measure the distribution of interaction strengths, even when the average value of the interaction is zero.
    By independently measuring distributions of different forms of disorder, this methodology can elucidate which microscopic symmetry drives a phase transition.
    We discuss potential applications to study complex  phase transitions in strongly interacting quantum materials.
\end{abstract}

\maketitle

\section{Introduction}

The study of symmetry-breaking phase transitions is a cornerstone of condensed matter physics.
Landau theory explains such transitions classically: symmetry-related degenerate ground states appear below a critical temperature, and the symmetry breaks as the system chooses one of these state over others~\cite{LandauBook}.
The loss of symmetry is often understood through the definition of an order parameter: a (possibly observable) quantity which describes the magnitude of the symmetry-breaking in the material and encodes some macroscopic property of the system.
Fluctuations of the order parameter around its zero value grow exponentially when approaching a critical temperature from above, making it an important indicator of the emerging phase~\cite{Lipa1996}.
In systems where the degenerate ground state manifolds are caused by frustration, e.g. ``accidental'' symmetries of the Hamiltonian, a different type of symmetry-breaking can occur.
In this case, one specific ground state becomes preferred due to differences in low-energy fluctuations, a process known as order by disorder (ObD)~\cite{Villain1980, Savary2012}.

\begin{figure}
    \centering
    \includegraphics[width=\linewidth]{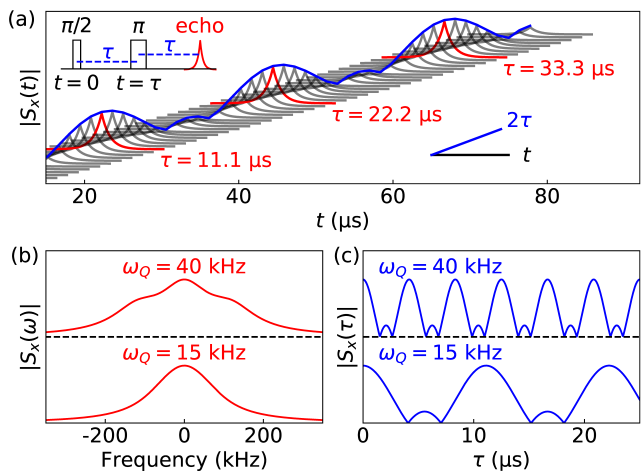}
     \vspace{-0.5cm}
    \caption{
    (a)~Two-dimensional spectroscopy for a spin-3/2 nuclei with quadrupolar interaction strength $\omega_Q = 15$~kHz (and no quadrupolar noise, $\Gamma_Q = 0$) and magnetic (Zeeman) noise $\Gamma_Z = 100$~kHz.
    Values of $\tau$ that give local maxima in the echo amplitude are highlighted in red.
    The upper inset shows the echo pulse sequence, and the lower inset shows the two independent time-parameters, experiment time $t$ and total integration time $2\tau$.
    (b)~Traditional echo spectra for $\omega_Q = 15$~kHz and 40~kHz. Note that for $\omega_Q = 15$ kHz, the splitting caused by the quadrupolar interaction is lost within the magnetic-noise ($\Gamma_Z$) dominated linewidth.
    (c)~Echo amplitudes as a function of integration time $\tau$ for the same two values of $\omega_Q$. }
    \label{fig:intro}
\end{figure}

To understand both traditional and ObD symmetry-breaking phase transitions, one must first understand the origins and size of the associated fluctuations above the critical temperature.
For example, consider super-linear spin interactions ($\S_z^n$, $n > 1$), which can be caused in both electronic and nuclear Hamiltonians by anisotropic electric field gradients or strong spin-orbit interactions.
The most commonly studied of these interactions is the quadrupolar type ($\S_z^2$), associated with an interaction strength $\omega_Q$, but order parameters of higher power terms are also possible~\cite{Sivardiere1973}.
Phases driven by octopolar ($\S_z^3$)~\cite{Maki1979,Bohmer1990,Kiss2005} and hexadecapolar ($\S_z^4$)~\cite{Toth2011,Senyuk2016} interactions have been examined in some materials, but we focus here on the more common quadrupolar interaction.

Almost all techniques for measuring quadrupolar order are only sensitive to the average atomic or electronic structure of the material~\cite{Caciuffo2003,Onimaru2005,Bombardi2008,Shen2019,Maharaj2020}.
For example, in neutron diffraction, the measurements of local structure requires high energy sources or extremely accurate scattering models (form factors) to access local information about the material~\cite{Proffen2003,Fornasini2015,Welberry2016,Pourovskii2021}.
Moreover, if multiple sources of disorder are present, e.g. lattice distortions, magnetic noise, and orbital fluctuations, disentangling them can be challenging as they all contribute to the cross section~\cite{Zaliznyak2015}.
This can leave lingering questions about the nature of the phase transition above the critical temperature.

Some of these shortcomings have been overcome in the few-spin case (molecular or spin-qubit systems), due to developments in quantum information technology and chemical spectroscopy techniques~\cite{Brown1997,Madi1998,Smallwood2018}. 
One such approach is multi-dimensional spectroscopy, originally developed for magnetic resonance~\cite{Muller1975}.
By taking Fourier transforms of experimental signals that depend on pulses occurring at multiple independent times, one obtains spectral information in a high-dimensional space (one dimension for each of the associated pulse or delay times), as shown in Fig.~\ref{fig:intro}(a).
The multi-dimensional approach has been leveraged in few-spin systems to extract different symmetry-specific terms in a Hamiltonian~\cite{Lemmer2015,Szakowski2017,Sung2021}, and to study extended dipolar-coupled spin networks~\cite{Cho2005,Cho2006}.
Our goal in this work is to translate these techniques, well known in quantum information and chemical spectroscopy, to the study of phase transitions in many body systems.

The introduced methodology distinguishes the noise from inversion symmetric ($\mathcal{S}_i^n$, $n$ even) and inversion asymmetric ($n$ odd) terms in microscopic Hamiltonians.
In most materials, inversion symmetric noise often originates from charge or orbital degrees of freedom, while inversion asymmetric noise arises from magnetic disorder.
This noise, which is defined as the distribution of the parameters of the single-spin Hamiltonian, can be caused by both spatial or temporal variations.
However, the temporal variations must occur on a time scale longer than the experiment time ($2\tau$) but much shorter than the total run-time of an integrated spectral acquisition (e.g. when the spectra is obtained from many repetitions of an identical experiment).
We showcase this methodology for the characterization of phase transitions, providing a non-intrusive technique that directly measures the \textit{distribution} of symmetry breaking terms and not just their mean value.

In this paper, we describe a specific two-dimensional spectroscopic technique for a spin-$3/2$ system with a quadrupolar interaction in Section \ref{sec:quad}, which explains sinusoidal oscillations of the echo amplitude in $\tau$.
In Section~\ref{sec:distributions}, we derive how these oscillations transform under distributions of magnetic and quadrupolar interaction strengths, and provide a $\tau$-spectroscopy method for the analysis of order parameter distributions.
We apply this method to a number of realistic experimental situations in Section~\ref{sec:examples}, and include a few notes on higher spin cases.
We summarize our results and discuss future applications and extensions in Section \ref{sec:conlucusion}.
Following the main text, we provide appendices which outline important, but tedious, calculations: the angle-dependence of the quadrupolar Hamiltonian in a rotating frame (App.~\ref{app:rotations}), the pulse-angle dependence of the echo magnetization (App.~\ref{app:general_theta}), and generalizations of the $\pi$-pulse result to higher spins (App.~\ref{app:halfspin} and App.~\ref{app:intspin}).

\section{Spin $> 1/2$ Echos}
\label{sec:quad}

% AND add to abstract
% Begin with discuss in context of NMR, but this methodology is applicable to any spin > 1/2 system, for magnetic-noise resistant measurements of electric  fields in solid-embedded atomic and molecular quantum sensors (e.g Ru?).
% Conclusion? Applications (Sec IV?)
% ESR doesnt work for measurement of multipolar order: either (1) the electron  is localized on an atom and therefore is not sensitive to the tensor structure of the order, or (2) the electrons are delocalized, but then the orbital structure prevents spin from being a good quantum number.

Although our proposed methodology is general for magnetic Hamiltonians of arbitrary power, we will focus on the simplest example: one that contains only a magnetic term ($\S_z$) and an inversion-symmetric one ($\S_z^2$).
We will work out the problem for nuclear spins in the context of NMR measurement of crystals.
However, this same methodology can be used for developing quantum sensing and control protocols of higher spin atomic or molecular qubits embedded in solid matrices. 

Nuclei with spin greater than $1/2$ can couple to electric and magnetic moments, making them a useful probe of multipolar electronic phases.
Compared to the pair of central spectral lines obtained from integer-spins, half-integer spins provide a stronger, single central line, making them more robust for experiments.
For these reasons, we will focus on a spin $3/2$ Hamiltonian in the presence of an electric field gradient (EFG), given by the Hamiltonian~\cite{Pound1950} (see App.~\ref{app:rotations}):
\begin{equation}
%    \H = \omega_Z \S_z + \frac{\omega_Q}{2} \left[ (3 \S_z^2 - \S^2) + \eta (\S_x^2 - \S_y^2) \right]
\H = \vec{\omega_Z} \cdot \vec{\S} + \vec{\S}^\dagger \cdot \vec{Q} \cdot \vec{\S}.
\end{equation}
The vector $\vec{\omega_Z}$ is a magnetic noise term, whose magnitude and direction may vary across a sample due to magnetic field inhomogeneity or disorder in the crystal.
For most experiments, a large field is applied in one direction (canonically identified as the $z$-axis) such that any out-of-plane magnetization of the nuclei precesses at a very fast rate about the applied field.
In other words, the experiment is performed in a rotating frame which is ``locked in'' to a given resonant frequency $\omega_0$, usually on the order of hundreds of MHz.
In this rotating frame, any in-plane components of $\vec{\omega_Z}$ are effectively averaged to zero, and the rank-2 tensor $\vec{Q}$ averages out to a highly simplified form (App.~\ref{app:rotations}), yielding an effective rotating-frame Hamiltonian of
\begin{equation}
    \label{eq:H}
    \tilde{\H} = \omega_Z S_z + \frac{3 \omega_Q}{2} S_z^2
\end{equation} 
where $\omega_Z$ now describes the local deviation from the chosen resonant frequency $\omega_0$, and $\omega_Q$ provides the local quadrupolar moment.
We note that the quadrupolar interaction strength commonly used in the NMR literature is given by $\nu_Q \equiv 3\omega_Q / (2 \pi)$.

An EFG is most commonly caused by a distortion of the crystal lattice which removes a rotational symmetry.
Near a phase transition, as the crystal transitions from the symmetric to distorted phase, the energies of the distorted symmetry-broken sites approach those of the symmetric sites.
Due to this near degeneracy, the distorted structures begin to appear throughout the material, with thermal fluctuations allowing them to appear for short periods of time in small domains.
The values of the order parameter which captures the strength and direction of these distortions, the tensor $\vec{Q}$, therefore acquires a broad distribution across the sample just before the critical temperature.
However, estimating the distribution of $\vec{Q}$ (or its scalar projection along a specified applied field axis, $\omega_Q$) just before the transition via traditional one-dimensional spectroscopic techniques is difficult: its average value is \textit{zero} so only broadening of the spectral line can give insight, but this information is often hidden inside the line broadening caused by magnetic noise (distributions of $\omega_Z$).
Since the Hamiltonian term associated with $\omega_Q$ is non-linear, the effect of the broadening on the acquired spectrum is not straightforward and may be difficult to evaluate and interpret if it is smaller than other forms of disorder in the system.

Existing multi-dimensional spectroscopy methods only extract the average value of the quadrupolar interaction~\cite{Antonijevic2005}, and do not provide a clear pathway to accessing the distribution of $\omega_Q$.
We will show that a careful analysis of Hahn echos of varying pulse separation times $\tau$ reveals \textit{both the strength and distribution of} $\omega_Q$.
We begin by reviewing the spin-echo dynamics for a nuclei in a quadrupolar field~\cite{Abe1966}.
Periodic dependence of the echo amplitudes on $\tau$ has been previously observed in quadrupolar systems~\cite{Abe1964, Abe1966, Vachon2006}, but it is much less common in the literature than the conventional one-dimensional ($t$-domain) spectroscopic techniques.
Our main result is how the echo-amplitude behaves under variations in the distribution of $\omega_Q$, which is not present in these previous works.

%The presence of even a small finite value of $\omega_Q$ in Eq.~\ref{eq:H} is associated with a symmetry-breaking phase, and our method provides an extremely sensitive probe of this parameter.
%This can be seen in a direct analysis of the dynamics of the Hamiltonian in Eq.~\ref{eq:H}.

\subsection{The spin echo ($\pi/2$-$\tau$-$\pi$-$\tau$)}

We now describe the spin echo ``experiment'',  with a pulse sequence notated as $\pi/2$-$\tau$-$\pi$-$\tau$, followed by detection.
It is performed by applying the following four operations upon the density matrix:
\begin{enumerate}
    \item Rotation into the $x-y$ plane by a $90^\circ$ ($\pi/2$) rotation about the $y$ axis, via application of the operator $\R_y(90^\circ) = e^{-i \pi I_y/2}$.
    \item Time evolution to time $\tau$, via application of the unitary time evolution operator $\U(\tau) = e^{-i \tilde{\H} \tau}$ (note that $\tilde{\H}$ is diagonal).
    \item Rotation about the $x$ axis by $180^\circ$ ($\pi$), to cancel any accumulated phase due to $\omega_Z$, via application of the operator $\R_x(180^\circ)  = e^{-i \pi I_x}$.
    \item Time evolution to time $2\tau$, via application of the same unitary time evolution operator $\U(\tau)$ as above.
\end{enumerate}
We have assumed that $\hbar = 1$ to simplify notation, and set the units of the Hamiltonian's parameters to either angular frequency or conventional frequency depending on the application.
Throughout this work, the magnetic field used to apply $\R_y$ and $\R_x$ is assumed to be much stronger than any other terms in $\tilde{\H}$.
In this case, we can treat the pulses as instantaneous and consider them as ideal spin-rotation matrices.
If instead the applied fields were comparable in strength to the terms in $\tilde{\H}$, one must instead consider $R_x = \exp[-(\tilde{\H} + \delta I_y)t]$ for an applied field of strength $\delta$.
This leads to more complicated, but still unitary, generalized rotation matrices (discussed further in App.~\ref{app:general_theta}).

We can calculate the final density matrix as a product of these operators:
\begin{equation}
\rho(2\tau) = \U(\tau) \R_x \U(\tau) \R_y \rho(0) \R_y^{-1} \U(\tau)^{-1} \R_x^{-1} \U(\tau)^{-1} 
\end{equation}
We assume the initial density matrix $\rho(0) \propto \S_z$, describing a mixed thermal state consisting of the various Zeeman (magnetic) energy levels $m$ in the applied field.
Tedious but straightforward calculations yield the following expression for the NMR signal (which is proportional to the $x$-component of the magnetization) at time $t=2\tau$:

\begin{equation}
\braket{\S_x (2\tau)} = \textrm{Tr}[\S_x \rho(2\tau)] \propto 1 + \frac{3}{2} \cos(6 \omega_Q \tau)
\label{eq:S_threehalf}
\end{equation}

This calculation of $\braket{S_x (2\tau)}$ predicts the magnetization perpendicular to the applied field in the rotating frame, and is captured in experiment by solenoid axes commonly used for inductive detection of NMR signals. 
This expressions predicts that the echo's peak amplitude is dependent on the ``integration time'' of the experiment, $\tau$, as shown in Fig.~\ref{fig:intro}(a).
We expect minima in the echo amplitude to occur when
\begin{equation}
    6\omega_Q \tau = \pi + 2\pi n
    \label{eq:supression_cond}
\end{equation}
and maxima when
\begin{equation}
    6\omega_Q \tau = 2\pi n
\end{equation}
for $n \in \mathbb{Z}$ (an integer).

\subsection{Generic pulse angles}

One does not need to be constrained to only the $\pi/2$-$\tau$-$\pi$ pulse sequence.
To further investigate the behavior of $\braket{S_x(2\tau)}$ as a function of the $\omega_Q$ interaction strength, we also consider signals obtained after a $\theta$-$\tau$-$2\theta$ pulse sequence, where $\theta$ is an arbritrary angle.
By relaxing the assumption of ideal pulsing, e.g. replacing $R_y(90^\circ)$ and $R_x(180^\circ)$ with arbitrary unitary matrices, an analytic expression for 
$\braket{\S_x (2\tau)}$ can be obtained~\cite{Abe1966}.
Note that the following expressions are still obtained under the strong-pulse approximation introduced earlier.
The full calculation is provided in App.~\ref{app:general_theta}, but it is identical in spirit to the calculation we just performed, with two small modifications.

First, the initial rotation matrix $\R_y$ takes an arbitrarily angle $\theta_1$, yielding a density matrix just after this first pulse of
\begin{equation}
\rho'(0) = (\cos \theta_1) \S_z + (\sin \theta_1) \S_x. 
\end{equation}
As the matrix multiplication and evaluation of the trace are linear operators, by first calculating the result for a density matrix proportional to $\S_z$ and one proportional to $\S_x$ independently, we can obtain the total value of $\braket{\S(2\tau)}$ without explicitly including $\theta_1$ in any matrix calculations.

\begin{figure}
    \centering
    \includegraphics[width=\linewidth]{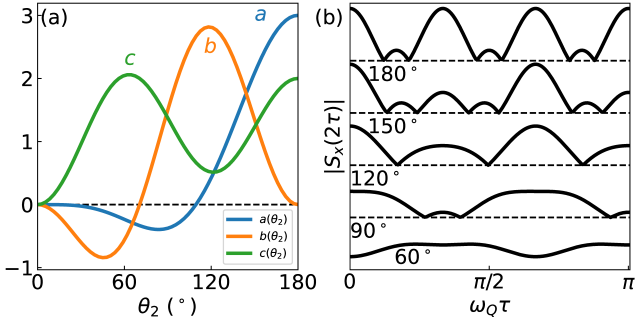}
    \caption{
    (a) The pulse-angle ($\theta_2$) dependence of the $\{a,b,c\}$ functions (Eq.~\ref{eq:S_coeffs}).
    (b) Value of $\braket{\S_x(2\tau)}$ as a function of the dimensionless parameter $\omega_Q \tau$ for five pulse-angles between $60^\circ$ and $180^\circ$.}
    \label{fig:theta_dependence}
\end{figure}

Second, we assume the $\R_x$ pulse is a generic (unitary) matrix, and rewrite its entries in terms of the second pulse angle $(\theta_2)$.
This lengthy algebraic process yields the following form, after removing terms that depend on deviation from the resonance frequency ($\omega_Z$):  

\begin{equation}
\label{eq:S_exact}
\begin{split}
\braket{\S_x (2\tau)} &\propto a(\theta_2) \cos(6 \omega_Q \tau) + b(\theta_2) \cos(3 \omega_Q \tau) + c(\theta_2).
\end{split}
\end{equation}
with
\begin{equation}
\label{eq:S_coeffs}
\begin{split}
a(x) &= -\frac{3}{2} (1 + 3 \cos x) \sin^4 \frac{x}{2}, \\
b(x) &=  \frac{3}{2} (1 - 3 \cos x) \sin^2 x, \\
c(x) &=  \frac{1}{8} (49 + 60 \cos x + 27 \cos 2x) \sin^2 \frac{x}{2}.
\end{split}
\end{equation}
One can easily check that for $\theta_2 = \pi$, we recover $a = 3$, $b = 0$, and $c = 2$, matching Eq.~\ref{eq:S_threehalf}.
We have plotted these three function in Fig.~\ref{fig:theta_dependence}(a), and examples of $\braket{\S_x (2\tau)}$ at different $\theta_2$ in Fig.~\ref{fig:theta_dependence}(b).

In Eq.~\ref{eq:S_exact} we have for now omitted terms which are proportional to $\cos(\omega_Z \tau)$ or $\sin(\omega_Z \tau)$, as their contribution is negligible in realistic cases of narrow magnetic noise ($\omega_Z$) distributions.
We will give a full discussion of their role in Sec.~\ref{sec:examples}, and the details of their calculation are given in App.~\ref{app:general_theta}.

\subsection{Sample tilt angle}

We also derive the corrections associated with a physical rotation of the sample such that the principle axis system (PAS) of the crystalline EFG no longer aligns with the applied field (App.~\ref{app:rotations}).
We work with Euler angles given by first a rotation about the (laboratory) $z$-axis by $\alpha$, then about the $x$-axis by $\beta$, and finally again about the $z$-axis by $\gamma$.
As we are working in the rotating frame, this last angle $\gamma$ is then averaged out via $\frac{1}{2\pi} \int_0^{2\pi} d\gamma$.
We find that the only change to the Hamiltonian (and resutling equations) is that the quadrupolar frequency $\omega_Q$ is replaced by an effective frequency due to the tilting of the sample:
\begin{equation}
\label{eq:tilt_angle}
\omega_Q^{\textrm{eff}} = \frac{\omega_Q}{2} \left( 3 \cos^2 \beta - 1 - \eta \cos 2 \alpha \sin^2 \beta \right)
\end{equation}
where $\eta$ is the anisotropy of the quadrupolar tensor in the PAS.

The magnetic term $\omega_Z$ can also have explicit tilt-angle dependence,  e.g. $\omega_Z(\alpha, \beta)$.
This can be caused by anisotropic shielding of the magnetic field by electronic structure, which is a well-known effect in solid-state magnetic resonance spectroscopy~\cite{Reif2021}, but will not be discussed here.

\section{Magnetic and quadrupolar distributions}
\label{sec:distributions}

In this section, we describe ways in which $\tau$-spectroscopy can be employed to independently probe magnetic and quadrupolar parameter distributions.
Instead of studying specific values of $\omega_Z$ and $\omega_Q$ for the single-spin Hamiltonian of Eq.~\ref{eq:H}, we will consider a large collection of independent spins with parameters following distributions $g_Z(\omega_Z)$ and $g_Q(\omega_Q)$.
This typically occurs when performing experiments on a micron-sized single-crystal sample, which has billions of nuclei each with their own local values of the two parameters.
It could also occur for a single spin experiment but with a specific scale of time-dependence in the parameters.
For this method to still apply, the value of $\omega_Q$ and $\omega_Z$ need to stay nearly constant during the $2 \tau$ duration of a single echo experiment, but vary between repeated experiments, common in experiments as averaging via repetition is needed to improve signal-to-noise ratios.

For a specific value of the two parameters, say $\omega_{Z_1}$ and $\omega_{Q_1}$, the response from that specific spin Hamiltonian will be assigned a weight $g_Z(\omega_{Z_1}) \times g_Q(\omega_{Q_1})$.
The total response of any spectral experiment will therefore be:
\begin{equation}
    F_\textrm{tot} = \int \int d \omega_Z d \omega_Q g_Z(\omega_Z) g_Q(\omega_Q) F(\omega_Z, \omega_Q)
\end{equation}
where $F(\omega_Z, \omega_Q)$ is some response from a specific, single-spin Hamiltonian (Eq.~\ref{eq:H}), and $F_\textrm{tot}$ is the aggregate response across the entire sample.

Before considering the effect of setting $F = \braket{S_x(2\tau)}$, we first examine how traditional time-domain spectra can fail to capture such distributions.

\subsection{Shortcomings of the traditional approach}

We directly calculate the spectral response, $F = A(\nu)$, under a $\pi/2$-$\tau$-$\pi$ experiment due to the Hamiltonian of Eq.~\ref{eq:H}.
Assuming an initial thermal ensemble, $\rho \propto S_z$, $A(\nu)$ is defined as the Fourier transform of $\textrm{Tr}[S_x \rho(t)]$, which is given by:
\begin{equation}
    \begin{split}
    A(\nu; \omega_Q, \omega_Z) &\propto \int e^{- 2 \pi i \nu t} \textrm{Tr}[S_x e^{-i \tilde{\H} t} S_x e^{i \tilde{\H} t}] \\
    &= \tfrac{1}{10}[4 \delta(\nu - \tfrac{\omega_Z}{2 \pi}) + 3 \delta(\nu - \tfrac{\omega_Z + 3 \omega_Q}{2 \pi}) \\
    & + 3 \delta(\nu - \tfrac{\omega_Z - 3 \omega_Q}{2 \pi})], 
    \end{split}
\end{equation}
where we have explicitly written the $\omega_Q$ and $\omega_Z$ dependence into $A$, as they appear in the rotating frame Hamiltonian, $\tilde{\H}$.
Note that the argument ($\nu$) is given in units of conventional frequency, Hz, while our Hamiltonian parameters ($\omega_Z, \omega_Q$) are technically given in units of angular frequency, radians per second.
Note however that in all relevant figures that consider specific values of $\omega_i$ or their distributions ($\Gamma_i$), we provide their values in Hz, so their value in angular frequency (for use in Eq.~\ref{eq:S_threehalf}, for example) is given by multiplying this value by $2 \pi$.

First, let's consider a material with magnetic noise given by the distribution $g_Z(\omega_Z)$, but no quadrupolar noise ($g_Q(\omega_Q) = \delta(\omega_Q - \omega_{Q_0})$).
For simplicity, assume the distribution $g_Z$ is Lorentzian with linewidth $\Gamma_Z$ centered at $\omega_Z = 0$ (because of the rotating frame approximation).
The spectrum obtained is given by:
\begin{equation}
    \begin{split}
    A(\nu) &= \int d\omega_Q \delta(\omega_Q - \omega_{Q_0}) \int d \omega_Z g_Z(\omega_Z) A(\nu; \omega_Z, \omega_Q) \\
    &\propto 4 g_Z(2 \pi \nu) + 3 g_Z(2 \pi \nu \pm 3 \omega_Q).
    \end{split}
\end{equation}

\begin{figure}
    \centering
    \includegraphics[width=\linewidth]{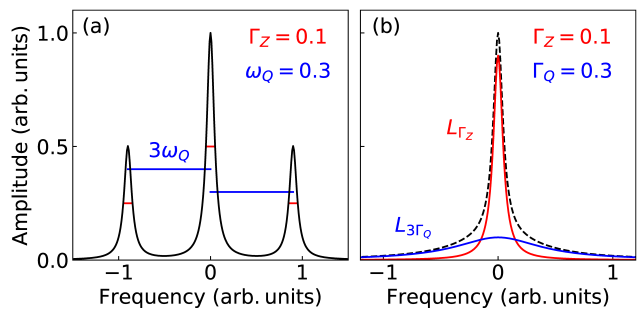}
    \caption{
    (a) Spectral transform of a spin echo with both Zeeman noise ($\Gamma_Z$, red) and a quadrupolar interaction ($\omega_Q$, blue).
    The location of the satellite peaks are  $\pm 3 \omega_Q$.
    (b) Spectral transform of a spin echo with Zeeman and quadrupolar noise, both centered at zero. The total spectra (black dashed line) can be approximated as a sum of two Lorentzians, $L_\Gamma$, with different FWHM values $\Gamma$.}
    \label{fig:z_vs_q}
\end{figure}

The quadrupolar term splits the spectrum into three peaks, a central transition line flanked by a pair of satellite peaks, both broadened by the magnetic noise $\frac{\Gamma_Z}{2\pi}$, as shown in Fig.~\ref{fig:z_vs_q}(a).
Note that even in the absence of quadrupolar noise, when $\Gamma_Z > \omega_Q$, identifying the location of the satellite peaks can be challenging, as was shown in Fig.~\ref{fig:intro}(b).

We next consider a sample where $\omega_Q$ also varies in space, and assume $g_Q(\omega_Q)$ follows a Lorentzian distribution centered at $0$ with linewidth $\Gamma_Q$.
This leads to an expression:
\begin{equation}
    A(\nu) \propto 4 g_Z(2 \pi \nu) + 3 \int d \omega_Q g_Q(\omega_Q) g_Z(2 \pi \nu \pm 3 \omega_Q).
\end{equation}
The second term are convolutions of the two noise distributions, one for the satellite peak above the central transition, and one for the peak below.
If we work under the assumption that $\Gamma_Z \ll \Gamma_Q$, we can approximate $g_Z(2 \pi \nu \pm 3 \omega_Q)$ as $\delta(2 \pi \nu \pm 3 \omega_Q)$ and obtain:
\begin{equation}
    A(\nu) \sim 4 g_Z(2 \pi \nu) + 9 g_Q(2 \pi \nu/3).
\end{equation}

This case is plotted in Fig.~\ref{fig:z_vs_q}(b), and we see that the clear three-peak signature of the quadrupolar interaction is lost.
As the central peak of the quadrupolar interaction is independent of $\omega_Q$, the variation in $\omega_Q$ is invisible to that proportion of the spectral weight ($40\%$).
This leads to a large central peak whose width is given by $\frac{\Gamma_Z}{2\pi}$.
The other portion of the spectral weight (corresponding to the two satellite peaks of the quadrupolar interaction) forms a distribution whose width is given by $\frac{3 \Gamma_Q}{2 \pi}$.
The spectra is therefore a non-Lorentzian distribution whose FWHM is slightly larger than $\Gamma_Z$.
Importantly, when $3 \Gamma_Q \gg \Gamma_Z$, the broad nature of $g_Q$ relative to $g_Z$ makes it difficult to resolve this distribution of $\omega_Q$ through the conventional spectral technique, motivating the need for an improved method.
We now present such a method.

\subsection{Effect of $\omega$-distributions on echo amplitudes}

A main goal in this work is to extract the distributions of $\omega_Q$ and $\omega_Z$ simultaneously.
We begin this endeavour by setting $A = \braket{\S_x(2 \tau)} \equiv K(\tau; \omega_Z, \omega_K)$ where we have used this labelling as $K$ is a linear kernel under the $g$-transform.
The average $\tau$-dependent echo intensity is given by:
\begin{equation}
    I(\tau) = \int \int d\omega_Q d\omega_Z g_Z(\omega_Z) g_Q(\omega_Q) K(\tau; \omega_Z, \omega_Q, \tau).
\end{equation}
We find that for all assumptions of general spin and pulse angle (see App.~\ref{app:general_theta},~\ref{app:halfspin},~\ref{app:intspin}), $K$ consists of only linear combinations (products and/or sums) of terms of the form $A_i \cos(n_i \omega_i \tau) \equiv K_i$.
In contrast, the time-dependent response in the previous subsection yielded terms of the form $\delta(\omega_Q + \omega_Z)$, breaking the assumption of a linear transformation and leading to convolutions of $g_Z$ with $g_Q$.
Because of this linearity in the order parameters $\omega_Z$ and $\omega_Q$, we can always perform the $g_Z$ or $g_Q$ transforms independently, and thus only need to know the result of
\begin{equation}
    \label{eq:linear_kernel}
    I_i(\tau) = \int g_i(\omega_i) A_i \cos(n_i \omega_i \tau)
\end{equation}
to derive an expression for any complicated form of $K$.
To clarify the notation used in this sum, the index $i$ accounts for terms of different frequency scales ($n_i$), with contributions from both of the Hamiltonian parameters ($\omega_i$ is equal to $\omega_Z$ or $\omega_Q$).
For example, in the $S=3/2$ case given in Eq.~\ref{eq:S_threehalf}, we have $K_1 = 1$ and $K_2 = \frac{3}{2} \cos(6 \omega_Q \tau)$, and we can read off that $n_1 = 0$, $n_2 = 6$, and $\omega_1 = \omega_2 = \omega_Q$.

 To calculate here the transform of this generalized function $K_i = A_i \cos (n_i \omega_i \tau)$, an assumption of the form of $g_i$ is needed.
 If we take the $\omega_i$ to be normally distributed with some mean $\omega_{i0}$ and standard deviation $\sigma_i$, we obtain
\begin{equation}
\begin{split}
\label{eq:gauss_I}
    I_i(\tau) &= \int \frac{1}{ \sigma_i \sqrt{2\pi}} e^{-\frac{1}{2} \left( \frac{\omega_i - \omega_{i0}}{\sigma_i} \right)^2} A \cos(n \omega_i \tau) d\omega_i \\
    &= A e^{-\frac{1}{2}(n \tau \sigma_i)^2} \cos (n \omega_{i0} \tau).
\end{split}
\end{equation}

We observe that the oscillations in $\tau$ still have a characteristic frequency $\omega_{i0}$ but decay like the inverse of the distribution of frequencies, $(n\sigma_i)^{-1}$.

Now consider a Lorentzian distribution of frequencies, $g(\omega_i) = \frac{1}{2 \pi} \frac{\Gamma_i}{(\omega_i - \omega_{i0})^2 + (\Gamma_i/2)^2}$. Then we obtain:
\begin{equation}
\begin{split}
    \label{eq:lorentz_I}
    I_i(\tau) &= \int \frac{1}{2 \pi} \frac{\Gamma_i}{(\omega_i - \omega_{i0})^2 + (\Gamma_i/2)^2} A \cos(n \omega_i \tau) d\omega_i \\
    &= A e^{-(n/2) \Gamma_i \tau} \cos(n \omega_{i0} \tau).
    \end{split}
\end{equation}

A general expression for our spectroscopy method is obtained by applying these transforms to the general echo amplitudes derived in App.~\ref{app:general_theta}, for distributions of $\omega_Z$ and $\omega_Q$ simultaneously.
Assuming a Lorentzian distribution of linewidth $\Gamma_Z$ with central frequency $0$ for $\omega_Z$, and one of linewidth $\Gamma_Q$ and center $\omega_{Q_0}$ for $\omega_Q$, the final result is:
\begin{equation}
\begin{split}
\label{eq:tau_spectroscopy_formula}
\braket{\S_x (2\tau)} &= (a(\theta_2) + a(\theta_2 + \pi) e^{-\Gamma_Z \tau}) e^{-3 \Gamma_Q \tau} \cos (6 \omega_{Q_0} \tau) \\
& + (b(\theta_2) +  b(\theta_2 + \pi) e^{-\Gamma_Z \tau/2}) e^{-3 \Gamma_Q \tau/2} \cos (3 \omega_{Q_0} \tau) \\
& + (c(\theta_2) + c(\theta_2 + \pi) e^{-\Gamma_Z \tau}),
\end{split}
\end{equation}
with $\{a,b,c\}$ as given in Eq.~\ref{eq:S_coeffs}.
The assumption that the central frequency for magnetic fluctuations is zero, $(\omega_{Z0} = 0)$, is equivalent to assuming that the experiment is performed in the ``ideal'' rotating frame of the sample.

By looking at the $\tau$-dependence of the spin-echo amplitude, one can extract the quadrupolar linewidth $\Gamma_Q$ and the central frequency $\omega_{Q_0}$.
Furthermore, by changing the pulse duration (and thus the pulse-angle $\theta_2$), one can verify the variations in the $\{a,b,c\}$ coefficients, allowing us to confirm that the refocusing is indeed caused by a $\S_z^2$ term in the Hamiltonian.
In the following section, we will examine how this methodology would play out in a few realistic experiments.

\section{Applications of $\tau$ spectroscopy}
\label{sec:examples}

Here we illustrate the utility of $\tau$-spectroscopy in the identification of critical fluctuations of a inversion symmetric order parameter.
Large changes in the width of the $\omega_Q$ distribution ($\Gamma_Q$) is expected if the quadrupolar order is associated with a phase transition, and we will be able to distinguish these fluctuations from those of magnetic origin ($\Gamma_Z$).

\subsection{Temperature-dependent phase transition in $\omega_Q$}

For the simplest example of how to use Eq.~\ref{eq:tau_spectroscopy_formula}, consider an experiment which sweeps $\tau$ under the perfect pulsing condition ($\theta_2 = \pi$).
In this case, $a(\theta_2) = 3$, $b(\theta_2) = 0$, $c(\theta_2) = 2$.
The terms evaluated at $x = \theta_2 + \pi = 2 \pi \sim 0$ all evaluate to zero because a factor of $\sin(x)$ is present in each coefficient, and so we have no dependence on $\Gamma_Z$.

To replicate an experimental spectra, we also wish to consider a $T_2$ decay process.
This decay is caused by magnetic scattering between nuclei or to the electronic environment, and can be captured numerically by inclusion of non-unitary ``jump'' matrices in the Lindbladian master equation for open quantum systems.
For our purposes, we assume this $T_2$ process is slower than any dephasing caused by the distributions of $\omega_Z$ and $\omega_Q$, and can thus be included as an overall factor to the large-$\tau$ steady-state result by including a factor $e^{-2\tau/T_2}$ (the factor of $2$ is due to measuring $T_2$ in terms of of experiment time, e.g at $t = 2\tau$).
This yields a $\tau$-dependent spin-echo amplitude $I$ of:
\begin{equation}
\label{eq:I_pi}
    I(\tau) = \frac{I(0)}{5} \left(3 e^{-3 \Gamma_Q \tau}\cos 6 \omega_{Q_0} \tau  + 2\right) e^{-2\tau/T_2}
\end{equation}
where $I(0)$ is the spin-echo amplitude at $\tau = 0$ (e.g. the inital magnetization).

\begin{figure}
    \centering
    \includegraphics[width=\linewidth]{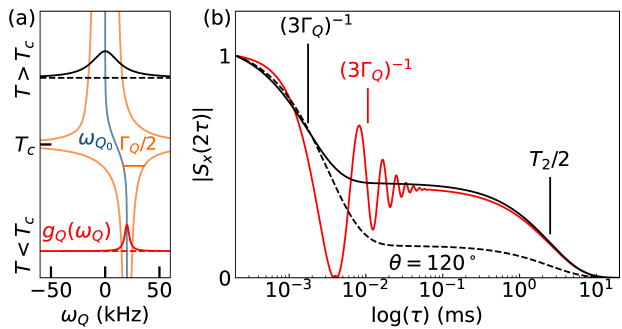}
    \caption{
    (a) Temperature dependence of the distribution $g_Q(\omega_Q$).
    The central frequency, $\omega_{Q_0}(T)$, is plotted in blue while the FWHM, $\Gamma_Q(T)$, is given by the span of the two orange lines.
    For a temperature $T > T_c$ ($T < T_c$), the central frequency is $\omega_{Q_0} = 0$ kHz (20 kHz) and the linewidth is $\Gamma_Q = 30$ kHz (5 kHz), plotted in black (red).
    (b) The resulting $\tau$-dependence of the echo amplitude at the two chosen temperatures, with the quadrupolar dephasing time ($(3 \Gamma_Q)^{-1}$) and magnetic decoherence time ($T_2/2$) noted. Here, $T_2 = 5$ ms. The $\tau$-dependence for the black distribution at a non-ideal pulsing of $\theta_2 = 120^\circ$ is plotted in the black dashed line.
    %The resulting plateaus (the relative intensity at the intermediate time $\tau = 1$ ms) are indicated on the left-hand side.
    }
    \label{fig:omega_distrib}
\end{figure}

Now consider that at some temperature ($T$) above a critical temperature, $T > T_c$ in a given material, the $\omega_Q$ values are zero on average (Fig.~\ref{fig:omega_distrib}(a)).
However, there is a distribution of non-zero $\omega_Q$ values due to small EFG's caused by thermal fluctuations in the atomic or electronic structure.
Now imagine that for $T < T_c$, the system undergoes a transition such that there is an average finite EFG everywhere, with a globally aligned principal axis system (PAS).
Now the distribution of $\omega_Q$ has non-zero center, $\omega_{Q_0} \neq 0$.

These two cases of $\omega_Q$ distributions are plotted in Fig.~\ref{fig:omega_distrib}a, with the associated $\tau$-spectroscopy experiment shown in Fig.~\ref{fig:omega_distrib}b.
Importantly, even for $\omega_{Q_0} = 0$, \textit{the width of the distribution can easily be recovered from the $\tau$ spectroscopy experiment}.
This is because a ``plateau'' is observed between the $1/(3 \Gamma_Q$) and the $T_2/2$ decay time scales.
In other words, the presence of both orbital, or charge, noise and magnetic decoherence leads to a two-step relaxation process.
For $\omega_{Q_0} \neq 0$ sinusoidal variations occur, which clearly distinguish it from the $\omega_{Q_0} = 0$ case.
Note that the plateau for the $\omega_{Q_0} \neq 0$ case occurs in the upper-branch of the envelope of the decaying oscillations.
Most importantly, as $T \to T_C$ from either side, the exponential growth in $\Gamma_Q$ should be observable by a large reduction in the effective time scale $(3\Gamma_Q)^{-1}$ at which the plateau occurs.

We note that a dipole-dipole interaction between two isolated spin-$1/2$ particles can cause a similar two-step decay process. This is because an $S_z \otimes S_z$ interaction admits a term with similar form to the spin-$3/2$ $S_z^2$ term, and is inversion symmetric just like the quadrupolar interaction studied here.

\subsection{Verifying the pulse-angle dependence}

The two-step decay (plateau) in the $\tau$-dependent response is convincing evidence of a zero-centered order parameter distribution, but one may want to confirm that it is caused by the $S_z^2$ term of the Hamiltonian.
To do so, one can check that the prefactors $\{a,b,c\}$ in Eq.~\ref{eq:tau_spectroscopy_formula} behave as expected under longer or shorter pulses.
That is to say, bigger or smaller pulse-angles $\theta_2$ will modify the refocusing of the quadrupolar interaction in a predictable way.
After removing the assumption of perfect pulsing, there is also a correction to each of the three prefactors that will depend on the magnetic disorder, $\Gamma_Z$.
We therefore consider three distinct cases: $\Gamma_Z \gg \Gamma_Q$, $\Gamma_Z \ll \Gamma_Q$, and $\Gamma_Z \sim \Gamma_Q$.
In all three cases, we will fit our $\tau$-dependent echo intensity to the following functional form:
\begin{equation}
I(\tau) \sim \tilde{a}(\theta_2) e^{-3 \Gamma_Q \tau} + \tilde{b}(\theta_2) e^{-3 \Gamma_Q \tau /2} + \tilde{c}(\theta_2).
\end{equation}
Compared to Eq.~\ref{eq:I_pi} there is an additional term which decays half as fast (due to the $\cos 3 \omega_Q \tau$ term).
Fitting to this functional form consists of two steps.
First, estimating the quadrupolar dephasing time-scale from the $\tau$-value at which the plateau occurs, and then extracting the effective height and curvature near the plateau to estimate the prefactors $\{\tilde{a},\tilde{b},\tilde{c}\}$.

\begin{figure}
    \centering
    \includegraphics[width=\linewidth]{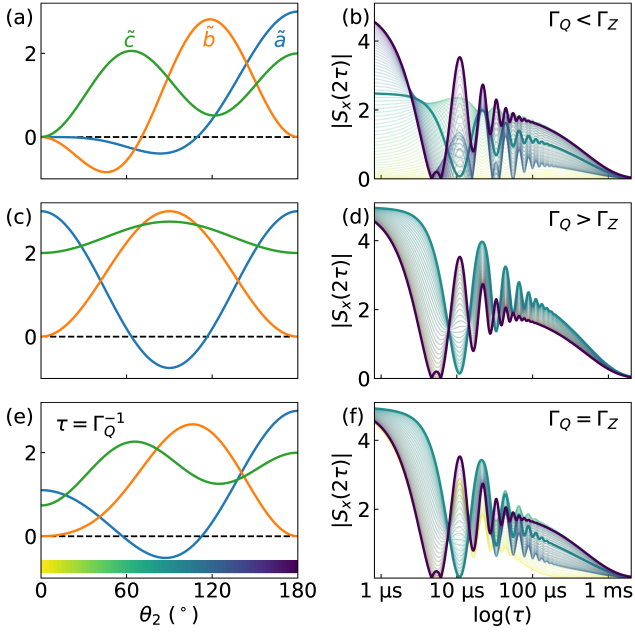}
    \caption{
    Dependence of the effective signal amplitudes $\{\tilde{a},\tilde{b},\tilde{c}\}$ on the pulse angle $\theta_2$ for (a) $\Gamma_Q < \Gamma_Z$, (b) $\Gamma_Q > \Gamma_Z$, and (c) $\Gamma_Q = \Gamma_Z$ (this last case is evaluated at integration time $\tau = \Gamma_Q^{-1}$).
    (b-f) The respective $\tau$-dependence of the echo amplitude for different $\theta_2$ (colored curves) for the three cases given in (a-c), indexed by the colormap at the bottom of (e). The curves at $\theta_2 = 90^\circ$ and $\theta_2 = 180^\circ$ are highlighted in teal and dark purple, respectively.
    For all simulations here, $\omega_{Q_0} = 15$ kHz, $\Gamma_Q = 3$ kHz, and $T_2 = 1$ ms.
    }
    \label{fig:pulse_dependence}
\end{figure}

In the first case, $\Gamma_Z \gg \Gamma_Q$, the magnetic dephasing is so fast that the second terms in the prefactors of Eq.~\ref{eq:tau_spectroscopy_formula} can be safely ignored and $\tilde{a}(\theta_2) \equiv a(\theta_2)$, as displayed in Fig.~\ref{fig:pulse_dependence}(a,b).
In the second case, $\Gamma_Z \ll \Gamma_Q$, the magnetic dephasing is so slow we can simply add the two terms together ($\tilde{a}(\theta_2) \equiv a(\theta_2) + a(\theta_2 + \pi)$), as displayed in Fig.~\ref{fig:pulse_dependence}(c,d).
However, for the third case, $\Gamma_Z \sim \Gamma_Q$, the spectroscopy becomes a bit more challenging.
Now there are $\Gamma_Q$-dependent and $\Gamma_Z$-dependent dephasings occurring simultaneously.
Thankfully, we can use the ``perfect'' pulsing condition (which refocuses $\Gamma_Z$ exactly) to estimate $\Gamma_Q$, and evaluation of the free-induction decay (FID) at $\theta_2 = 0$ to estimate $\Gamma_Z$.
Then, we can thoroughly understand the pulse-angle dependence as a smooth transition from the small $\Gamma_Z$ case (at small $\tau$) to the large $\Gamma_Z$ case (at large $\tau$), as seen in Fig.~\ref{fig:pulse_dependence}(e,f).

Perhaps the most important take away from Fig.~\ref{fig:pulse_dependence} is that the maximum amplitude at finite $\tau$ does not necessarily occur for the ideal pulsing, $\theta_2 = 180^\circ$.
That is to say, there are regions of $\tau$ in Fig.~\ref{fig:pulse_dependence}(b,d,f) where the teal curve ($\theta_2 < 120^\circ$) is larger than the purple curve ($\theta_2 = 180^\circ$).
In magnetic resonance experiments, the $\tau$ value is usually fixed during an initial pulse optimization sweep.
But without careful analysis of the pulse-dependent $\tau$-spectroscopy, it is impossible to be certain that one has not erroneously optimized the experiment at a $\tau$ value which attenuates the signal at $\theta_2 = 180^\circ$.
Thankfully, if $\omega_{Q_0} = 0$ there are no oscillations in $\tau$, and this attenuation problem will not occur.

\subsection{Higher spins}

We have seen that careful analysis of the $\tau$-dependence of a spin echo amplitude can give information about the distribution of $\omega_Q$ in a quadrupolar Hamiltonian.
We now generalize this technique to any spin and provide the formulae for the $\tau$-dependent responses under a $\pi/2$-$\pi$ echo sequence for spins up to $S = 11/2$.
The derivation of these formulae, which depend on iterative equations gained from a wavefunction treatment of the spin-echo problem, are provided in Appendix \ref{app:halfspin} and \ref{app:intspin}.

\begin{figure}
    \centering
    \includegraphics[width=\linewidth]{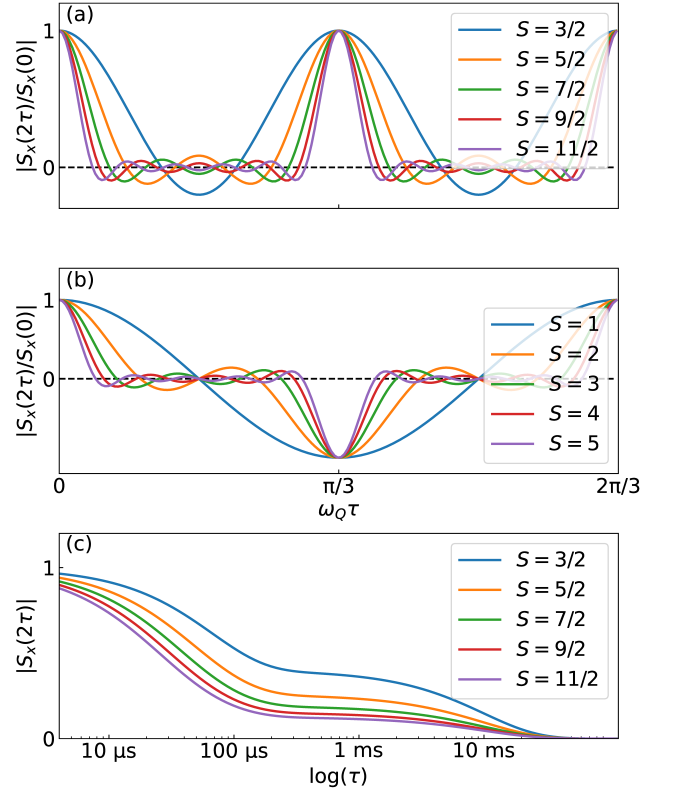}
    \caption{Dependence of $\braket{\S_x (2 \tau)}$ on $\omega_Q \tau$ for (a) half-integer spins and (b) integer spins. (c) $\tau$-dependent spectroscopy for half-integer spins, for a distribution with $\omega_{Q_0} = 0$ and $\Gamma_Q = 5$~kHz, and a magnetic decoherence time $T_2 = 10$~ms.}
    \label{fig:general_spin}
\end{figure}

The explicit forms up to $S = 5/2$ for half-integer spins are:
\begin{align}
    &\braket{S_x(2 \tau)} = \nonumber \\
        &\hspace{3mm}\begin{cases}
        2 + 3 \cos(6 \omega_Q \tau), &S=\tfrac{3}{2}. \\
        9 + 16\cos(6 \omega_Q \tau) + 10 \cos(12 \omega_Q \tau), &S=\tfrac{5}{2}.
        \end{cases}
\end{align}
In Fig.~\ref{fig:general_spin}(a) we plot the shape of each of these functions for $S = 3/2$ up to $S = 11/2$, with the equations for higher spin included in App. \ref{app:halfspin}.
For a distribution of $\omega_Q$, these higher frequency terms are acted upon by the linear transforms derived in Eq.~\ref{eq:gauss_I} and Eq.~\ref{eq:lorentz_I}.
The net effect of the additional, high frequency terms for the $\tau$-spectroscopy is a plateau that is both lower in amplitude and earlier in $\tau$, as shown in Fig.~\ref{fig:general_spin}(c).

The forms up to $S=2$ for integer spins are:
\begin{equation}
    \braket{S_x(2 \tau)} =
        \begin{cases}
        \cos(3 \omega_Q \tau), &S=1\\
        \cos(3 \omega_Q \tau) + 2\cos(9 \omega_Q \tau), &S=2
        \end{cases}
\end{equation}
and are plotted in Fig.~\ref{fig:general_spin}(b) up to $S=5$, with the equations for higher spin included in App.~\ref{app:intspin}.
As no constant term appears in these equations, a plateau between a $\Gamma_Q^{-1}$ timescale and the $\Gamma_Z^{-1}$ timescale will no longer occur under a distribution of $\omega_Q$.
A specially tailored pulse sequence beyond the simple Hahn echo is therefore needed to capture order parameter distributions in integer spin cases.

Although more cosine terms appear in the formula for $\braket{\S_x(2\tau)}$ as the spin number increases, the normalized values are converging to a simple function.
By inspection of the results at increasing (but finite) $S$, we obtain
\begin{equation}
\lim_{S \to \infty} \braket{\S_x (2\tau)} = \sum_{n=0}^\infty (-1)^{np} \delta (\omega_Q \tau - n \pi /3),
\end{equation}
where $p=2$ for half-integer spin, and $p=1$ for integer spin.

\section{Conclusion}
\label{sec:conlucusion}

We have developed a methodology for the determination of the mean-value ($\omega_0$) and distribution ($\Gamma$) of even-powered spin interactions ($\S_z^n$) in a solid state system, independent of all odd-powered interactions (Zeeman/magnetic).
We have focused on the quadrupolar Hamiltonian ($n = 2$), and provided closed form equations for general spins.
%in Appendices \ref{app:halfspin} and \ref{app:intspin}.
Even when the average value of $\omega_Q$ is zero, the effective linewidth of the distribution $\Gamma_Q$ is obtained from this straightforward $\tau$-spectroscopy if the quadrupolar timescale ($\Gamma_Q^{-1}$~or~$\sigma_Q^{-1}$) is smaller than the magnetic decoherence timescale ($T_2$).
Considering pulse-angle variations ($\theta_2 \neq \pi$), we find that the relative weighting of the $\tau$-dependent and $\tau$-independent terms change, which appears in experiments as a variable plateau height in the $\tau$-spectroscopy.
We also note that implementing an effective experiment for integer spin will be difficult as no $\tau$-independent term appears, suggesting a more sophisticated pulse-protocol should be developed in these cases.

Our methodology easily extracts the variations in the multipolar order parameter even when its average value is zero, giving crucial information about the temperature-dependent fluctuations that could drive a phase transition.
This provides invaluable physical insight into the mechanisms which drive magnetic frustration caused by the interplay of multiple interactions.
At the same time, the technique reveals any inversion or time-reversal symmetry among local interactions, giving direct evidence on the form of the microscopic Hamiltonian.

\begin{acknowledgments}
We thank M. Horvatić, D. E. Feldman and K. Plumb for helpful discussions.
This work was supported by the National Science Foundation under grant No. OIA-1921199. VFM also acknowledges support of the NSF grant No. DMR-1905532. A.D. was supported by the U.S. Department of Energy, Office of Science, Office of Basic Energy Sciences, under Award Number DE-SC0022311.

\end{acknowledgments}

\appendix

\section{Derivation of $H_Q(\alpha,\beta)$}
\label{app:rotations}
We define our quadrupolar frequency as:
\begin{equation}
    \omega_Q = \frac{e q V_{zz}}{S(2S-1)}
\end{equation}
where $V_{zz} \equiv eQ$ is the largest component of the EFG in the principle axis system (PAS, the basis choice where $V_{ij}~=~\partial E_i/\partial x_j$ is diagonal) and $eq$ is introduced to parameterize the nuclear coupling to the EFG.
In principle neither $eQ$ or $eq$ needs to be measured independently, as only their product enters into the observable $\omega_Q$.
We also define the assymmetry parameter as
\begin{equation}
    \eta = \frac{V_{xx} - V_{yy}}{V_{zz}}
\end{equation}
where again $V_{ii}$ is in the PAS.

We now derive how $H_Q$ changes if the material is rotated such that the laboratory frame does not align with the PAS.
Consider the material originally aligned with the PAS.
It then undergoes three rotations: first, a rotation about $z$-axis by $\alpha$,  then a rotation about the laboratory frame $x$-axis by $\beta$, and then again about the laboratory frame $z$-axis by $\gamma$.
The EFG ($V_{ij}$), which was originally diagonal in the PAS, now has components:
\begin{equation}
    \begin{split}
        V_{xx} = \frac{e Q}{2} &(3 \sin^2 \beta \sin^2 \gamma - \sin 2\alpha \cos \beta \sin 2 \gamma - 1 \\
        &+ \eta \left(\cos 2\alpha (\cos^2 \gamma - \cos^2 \beta \sin^2 \gamma) \right)), \\
        V_{yy} = \frac{e Q}{2} &(3 \sin^2 \beta \cos^2 \gamma + \sin 2\alpha \cos \beta \sin 2 \gamma - 1 \\
        &+ \eta \left(\cos 2\alpha (\sin^2 \gamma - \cos^2 \beta \cos^2 \gamma) \right)), \\
        V_{zz} = \frac{eQ}{2} &(3 \cos^2 \beta - 1 - \eta \cos 2 \alpha \sin^2 \beta), \\
        V_{xy} = \frac{eQ}{2} & ((3/2)(1 - \cos^2 \beta) \sin 2\gamma \\
        &- \eta ( (1/2)\cos 2\alpha (1 + \cos^2 \beta) \sin 2 \gamma\\
        &+ \sin 2 \alpha \cos \beta \cos 2 \gamma) ),\\
        V_{xz} = \frac{ eQ}{2} &(-(3/2)\sin 2\beta \sin \gamma \\
        &- \eta((1/2)\cos 2\alpha \sin 2\beta \sin \gamma \\
        &+ \sin 2 \alpha \sin \beta \cos \gamma),\\
        V_{yz} = \frac{eQ}{2} &(-(3/2)\sin 2\beta \cos \gamma \\
        &- \eta((1/2)\cos 2 \alpha \sin 2 \beta \cos \gamma \\
        &- \sin 2\alpha \sin\beta \sin \gamma)).
    \end{split}
\end{equation}
and $V$ is still symmetric and traceless.
As the spin echos are measured in the rotating frame, we can approximate (to first order in average Hamiltonian theory) the above terms by simply integrating $\gamma$ across the interval $[0,2\pi]$.
So in the rotating frame, all terms of $\cos \gamma$ or $\sin \gamma$ go to zero, and $\cos^2 \gamma$ and $\sin^2 \gamma$ go to $1/2$.
This yields the much simpler expressions:
\begin{equation}
    \begin{split}
     \tilde{V}_{zz} &= \frac{eQ}{2} (3 \cos^2 \beta - 1 - \eta \cos 2\alpha \sin^2 \beta), \\
     \tilde{V}_{xx} = \tilde{V}_{yy} &= \frac{eQ}{4} (3\sin^2 \beta - 2 + \eta \cos 2 \alpha (1 - \cos^2 \beta)), \\
     \tilde{V}_{ij} &= 0\textrm{, if } i \neq j.
    \end{split}
\end{equation}
The quadrupolar Hamiltonian for a spin in the presence of an EFG can be written via a dyadic inner product over spherical harmonics as~\cite{Pound1950}:
\begin{equation}
    H_Q = \vec{Q}^{(2)} \vec{\cdot} \vec{\nabla E}^{(2)} = \sum_{m=-2}^{2} (-1)^m Q^{(2)}_m \nabla E^{(2)}_{-m}.
\end{equation}
The nuclear spin terms are given by
\begin{equation}
    \begin{split}
    Q^{(2)}_0 &= A (3 \S_z^2 - \S^2), \\
    Q^{(2)}_{\pm 1} &= \mp A \sqrt{ \frac{3}{2}} (\S_{\pm} \S_z + \S_z \S_{\pm}),\\
    Q^{(2)}_{\pm 2} &= A \sqrt{ \frac{3}{2}} \S_{\pm}^2
    \end{split}
\end{equation}
with $A = eq/(2S(2S-1))$. The EFG components are
\begin{equation}
    \begin{split}
    \nabla E^{(2)}_0 &= \frac{1}{2} V_{zz}, \\
    \nabla E^{(2)}_{\pm 1} &= \mp \frac{1}{\sqrt{6}} (V_{xz} \pm i V_{yz}),\\
    \nabla E^{(2)}_{\pm 2} &= \frac{1}{\sqrt{6}} (V_{xx} - V_{yy} \pm 2i V_{xy}).
    \end{split}
\end{equation}

Only the first expression, $\nabla E^{(2)}_0$, is non-zero in the rotating frame.
By combining the dyadic expression for $H_Q$ with the basis transformation for $V_{ij}$, we finally obtain:
\begin{equation}
\begin{split}
    H_Q(\alpha,\beta) = \frac{3\omega_Q}{4} \S_z^2 \Bigl[3 \cos^2 \beta - 1 - \eta \cos 2 \alpha \sin^2 \beta \Bigr].
\end{split}
\end{equation}
Here we have dropped the irrelevant term $S^2$, as it is proportional to the identity.
This provides the quadrupolar part of the Hamiltonian in the PAS (second term of Eq. \ref{eq:H}) when $\alpha = \beta = 0$, and the general expression describes the $\omega_Q$ dependence on the sample tilt-angle (Eq.~\ref{eq:tilt_angle}).

\section{Derivation of $|\S_x(2 \tau)|$}
\label{app:general_theta}

Following Ref.~\onlinecite{Abe1966}, we will derive the full formula for $\braket{\S_x(2\tau)}$ for an arbitrary second pulse $U'$.
We assume the initial pulse $U$ transforms the density matrix into a form which can be written as a linear combination of $\S_z$ and $\S_x$.
Assuming $U'$ is given by a rotation about the $x$-axis, it takes the form
\begin{equation}
U'_x = \begin{pmatrix}
A & C & E & F \\
C & B & D & E \\ 
E & D & B & C \\
F & E & C & A
\end{pmatrix}.
\end{equation}
If instead $U'$ is given by a rotation about the $y$-axis,
\begin{equation}
U'_y = \begin{pmatrix}
A & C & E & F \\
-C & B & D & E \\ 
E & -D & B & C \\
-F & E & -C & A
\end{pmatrix}.
\end{equation}
Note that arbitrary $U'$ can be handled in a similar way, but then the relative angle between the generic complex coefficients $\{A,B,C,D,E,F\}$ must be taken into account, leading to a more complicated derivation~\cite{Abe1966}.
As the Hamiltonian in Eq.~\ref{eq:H} is diagonal, the time propagation by $\tau$ is given by a diagonal matrix whose entries are exponentials of the Hamiltonian's eigenvalues times $i\tau$.
We then calculate the contribution to $\textrm{Tr}(\S_x \rho(2\tau))$ for the two terms of the density matrix (one proportional to $\S_z$, the other to $\S_x$). 
We obtain

\begin{equation}
\begin{split}
S_x^{(z)} &=  f(\omega_Z \tau) \bigl( \sqrt{3} \left( BC + \sigma DE - 3 (AC + EF) \right) \\
&\times \cos 3 \omega_Q \tau - (6 CE - 2 \sigma BD) \bigr), \\ 
S_x^{(x)} &= -3C^2 - 2D^2 + 4\sqrt{3}BE \cos 3 \omega_Q \tau \\
&- 3 \sigma DF \cos 6\omega_Q \tau + (2B^2 + 3E^2) \cos 2 \omega_Z \tau\\
&- 4 \sqrt{3} \sigma CD \cos \omega_Z \tau \cos 3 \omega_Q \tau \\
&+ 3 AB \cos 2 \omega_Z \tau \cos 6 \omega_Q \tau.
\end{split}
\end{equation}
For an $x$-axis rotation at time $\tau$, $\sigma = 1$ and $f(x) = -i \sin x$.
For a  $y$-axis rotation at time $\tau$, $\sigma = -1$ and $f(x) = \cos x$.

The fact that only $3 \omega_Q \tau$ and $6 \omega_Q \tau$ appear in the arguments of the cosines is due to the fact that only specific sums or differences in the eigenvalues of the Hamiltonian appear in the final expression of the trace.
In particular, if we label the four eigenfrequencies of the diagonal Hamiltonian as $\omega_1$ through $\omega_4$, the first application of the $S_x$-like rotation leads to single-mixing terms of $\omega_{ij} = \omega_i - \omega_j$, with $\omega_{12} = \omega_Z + 3 \omega_Q$, $\omega_{23} = \omega_Z$ and $\omega_{34} = \omega_Z - 3 \omega_Q$.
And after the second pulse,  the frequency mixing leads to terms of the form $\omega_{12} - \omega_{34} \propto 6 \omega_Q$ and $\omega_{12} - \omega_{23} \propto 3 \omega_Q$. 
These mixings follow the well-known rule of magnetic resonance that only frequencies corresponding to certain pulse-induced energy-differences, or ``transitions'', appear in associated spectra.

Assuming $U'$ is a rotation about the $x$ or $y$-axis of the rotating frame by an angle $\theta_2$), one has:
\begin{equation}
\begin{split}
A &= \frac{1}{4} \left(3 \cos \frac{\theta_2}{2} + \cos \frac{3\theta_2}{2} \right), \\
B &= \frac{1}{4} \left(\cos \frac{\theta_2}{2} + 3 \cos \frac{3\theta_2}{2} \right), \\
C &= \frac{-s \sqrt{3}}{4} \left(\sin \frac{\theta_2}{2} + \sin \frac{3\theta_2}{2} \right), \\
D &= \frac{s}{4} \left(\sin \frac{\theta_2}{2} - 3 \sin \frac{3\theta_2}{2} \right), \\
E &= \frac{s^2\sqrt{3}}{4} \left(\cos \frac{\theta_2}{2} - \cos \frac{3\theta_2}{2} \right), \\
F &= \frac{-s^3}{4} \left(3 \sin \frac{\theta_2}{2} - \sin \frac{3\theta_2}{2} \right), \\
\end{split}
\end{equation}
where $s=i$ for an $x$-axis rotation and $s=1$ for a $y$-axis rotation.

The full expression for the pulse-angle dependence of the echo response at time $2\tau$ simplifies to
\begin{equation}
\begin{split}
S_x^{(x)} &= (\sigma a(\theta_2) + a(\theta_2 + \pi) \cos 2 \omega_Z \tau) \cos 6 \omega_Q \tau \\
& + (\sigma b(\theta_2) +  b(\theta_2 + \pi) \cos \omega_Z \tau) \cos 3 \omega_Q \tau \\
& + (\sigma c(\theta_2) + c(\theta_2 + \pi) \cos 2 \omega_Z \tau), \\
a(x) &= -\frac{3}{2} (1 + 3 \cos x) \sin^4 \frac{x}{2}, \\
b(x) &=  \frac{3}{2} (1 - 3 \cos x) \sin^2 x, \\
c(x) &=  \frac{1}{8} (49 + 60 \cos x + 27 \cos 2x) \sin^2 \frac{x}{2}.
\end{split}
\end{equation}
One can check that at the perfect pulsing condition, $\theta_2 = \pi$, all terms involving $\omega_Z$ drop out completely, as a perfect $\pi$-pulse refocuses all magnetic noise at $2 \tau$.

As the terms $\{a,b,c\}$ are symmetric about $\theta_2 = 0$ and $\theta_2 = \pi$, there is an ability to interchange $\omega_Z$-dependent terms with $\omega_Z$-independent terms by inverting $\theta_2$ about $\pi/2$. 
This symmetry captures the fact that one can robustly compare the free induction decay (the so-called FID, at $\theta_2 = 0$) to the ideal pulsing condition ($\theta_2 =\pi$) to verify the relative sizes of the quadrupolar and magnetic distribution linewidths, independent of the spin-decoherence time-scale $T_2$.

The other term that enters into total spin-echo magnitude, $S_x^{(z)}$, simplifies to:
\begin{equation}
S_x^{(z)} = h(\omega_Z \tau) \sin(\theta_2) \left(3 \cos 3 \omega_Q \tau - 2 \sigma \right)
\end{equation}
where $h(x) = \sin x$ for and $x$-axis rotation by $U'$, and $h(x) = \cos x$ for a $y$-axis rotation.
As $\sin \omega_Z \tau$ is odd, and distributions of $\omega_Z$ are often symmetric about the resonant frequency, this term goes to zero when averaging over the $\omega_Z$ distribution if $U'$ is an $x$-axis rotation.

\section{Half-integer spins}
\label{app:halfspin}

Generalizing the previous section to higher spin is straightforward, but tedious.
Here, we derive formula for $\braket{\S_x(2\tau)}$ at the $\theta_2 = \pi$ pulsing condition by way of iterative equations for each $m$ spin channel, instead of directly working with the density matrix.
This allows for a compact and scalable derivation of the $\tau$-spectroscopy to higher spins, albeit without the ability to assess the $\theta_2$ dependence.

We begin with a pedagogical review of the standard $SU(2)$ representation for general spin.
We will choose the $\S_z$ operator as the diagonal matrix with descending elements $\{S,S-1,\dots,-S+1,-S\}$. The operators $\S_x$ and $\S_y$ can be determined from the sum or difference of the related operators $\S_\pm = \S_x \pm i \S_y$. The momentum raising/lowering operators $\S_\pm$ are zero except for the terms given by
\begin{equation}
    \bra{S,j \pm 1} S_\pm \ket{S,j} = \sqrt{S(S+1) - j(j \pm 1)}.
\end{equation}
The general form of $\S_x$ and $\S_y$ in the $z$ basis are therefore
\begin{equation}
\begin{split}
    \S_x &= \frac{1}{2}\begin{pmatrix}
    0     &    a^S_S     &            &   \\
    a^S_S &   0         & a^S_{S-1}  &   \\
          &   a^S_{S-1} &      0     &   \\ 
          &             &            & \ddots
    \end{pmatrix}, \\
    \S_y &= \frac{i}{2}\begin{pmatrix}
    0     &    -a^S_S     &            &   \\
    a^S_S &   0         & -a^S_{S-1}  &   \\
          &   a^S_{S-1} &      0     &   \\ 
          &             &            & \ddots
    \end{pmatrix}
\end{split}
\end{equation}
with $a^S_j = \sqrt{S(S+1) - j(j-1)}$. Note that these matrices are symmetric about their center because $a^S_j = a^S_{-j+1}$.
To estimate the spin echo, we will need to know the initial state, which is given by a $90^\circ$ rotation about the $y$-axis from the $\braket{\S_z}=m$ state, $\ket{\psi_0}~=~\R_y^S(90^\circ) \ket{S,m}$.
We will also need the operator which performs a $180^\circ$ rotation about the $x$-axis.
Thankfully, the latter is quite simple in this basis
\begin{equation}
\R_x^S(180^\circ) = i^{2S+2} \begin{pmatrix}
  &   &   & 0 & 1 \\
  &   &   & 1 & 0 \\
  &   & $\reflectbox{$\ddots$}$ &   &   \\
0 & 1 &   &   & \\
1 & 0 &   &   &   
\end{pmatrix}
\end{equation}
but $\ket{\psi_0}$ must be determined from a set of iterative equations. The easiest way to obtain $\ket{\psi_0}=\sum_m \psi_m \ket{S,m}$ is to realize it must be an eigenvector of $\S_x$ with a specific eigenvalue $m$,

\begin{equation}
\label{eq:psi_0}
\frac{1}{2}\begin{pmatrix}
    0     &    a_S     &            &   \\
    a_S &   0         & a_{S-1}  &   \\
          &   a_{S-1} &      0     &   \\ 
          &             &            & \ddots
    \end{pmatrix}
    \begin{pmatrix}
    \psi_S \\
    \psi_{S-1} \\
    \psi_{S-2} \\
    \vdots \\
    \end{pmatrix}
    =
    m
    \begin{pmatrix}
    \psi_S \\
    \psi_{S-1} \\
    \psi_{S-2} \\
    \vdots \\
    \end{pmatrix}
\end{equation}
where we have begun to suppress the superscript $S$ for simplicity.
To derive the thermal state, we will consider a weighted sum of the results from different $m$ values.
Note that the choice of $m$ only enters into the overall calculation in the above equation and the resulting entries of $\ket{\psi_0}$.
To summarize the first three lines of Eq.~\ref{eq:psi_0}:
\begin{equation}
    \begin{split}
        a_S \psi_{S-1} &= 2m \psi_S, \\
        a_S \psi_S + a_{S-1}\psi_{S-2} &= 2m \psi_{S-1}, \\
        a_{S-1} \psi_{S-1} + a_{S-2}\psi_{S-3} &= 2m \psi_{S-2}. \\
    \end{split}
\end{equation}
Ignoring the overall normalization of $\ket{\psi_0}$ for now, we can set $\psi_S = 1$.
The first line then gives us $\psi_{S-1} = 2m/a_S$.
The rest of the $\psi_j$ are given by the recursion relationship
\begin{equation}
    \psi_j = \frac{1}{a_{j+1}} \left(2m \psi_{j+1} - a_{j+2} \psi_{j+2} \right)
\end{equation}
and note there is an inversion symmetry $\psi_j = \psi_{-j}$.
Now we assume our Hamiltonian is a diagonal matrix, $\H = \textrm{diag}\{h_S,h_{S-1},\dots,h_{-S}\}$ for general elements $h_j$. The corresponding time propagator is $U(\tau) = \textrm{diag}\{e^{-i h_S \tau}, e^{-i h_{S-1} \tau}, \dots , e^{-i h_{-S} \tau}\}$ and we can write the final (echo) state as
\begin{equation}
    \ket{2 \tau} = \U(\tau) \R_x(\pi) \U(\tau) \ket{\psi_0}.
\end{equation}
As $\R_x(\pi)$ simply inverts a state vector and multiplies it by $\pm i$,  we can quickly write down the final state as
\begin{align}
\ket{2 \tau} &= \textrm{diag}\left\{ i(-1)^{S+1/2} \psi_j e^{-i (h_j + h_{-j}) \tau} \right\} \\
&\equiv \textrm{diag}\{F_j\}. \nonumber
\end{align}
As $\psi_j = \psi_{-j}$, this state vector is symmetric about its center, and as the operator $\S_x$ is as well, we only need to evaluate half of the terms in inner product $\bra{2 \tau} \S_x \ket{2 \tau}$, e.g. from $j = S$ to $j = 1/2$. Taking special note of the structure caused by the off-diagonal terms of $\S_x$ (we have grouped the product terms from the inner product in brackets) we obtain
\begin{equation}
\begin{split}
    \frac{1}{2} \bra{2 \tau} \S_x \ket{2 \tau} = &\left[ a_S F_S^\dagger F_{S-1} \right] \\
    + &\left[ a_S F_{S-1}^\dagger F_S + a_{S-1} F_{S-1}^\dagger F_{S-2} \right] \\
    + &\left[ a_{S-1}F_{S-2}^\dagger F_{S-1} + a_{S-2}F_{S-2}^\dagger F_{S-3} \right] + \dots \\
    \dots + &\left[ a_{3/2} F_{1/2}^\dagger F_{3/2} + a_{1/2} F_{1/2}^\dagger F_{-1/2} \right]. 
    \end{split}
\end{equation}
Notice that by grouping across adjacent brackets, this becomes a sum of conjugate pairs except for an unpaired $j = 1/2$ term
\begin{equation}
\begin{split}
    \braket{\S_x (2\tau)} = &2\sum_{j = 1/2}^S a_j \psi_j \psi_{j-1} \\
    \times &\left[ e^{-i(h_j + h_{-j} - h_{j-1} - h_{-j+1}) \tau} + (\lambda_j-1) h.c. \right]
\end{split}
\end{equation}
with $\lambda_j = 1$ if $j = 1/2$, and $\lambda_j = 2$ otherwise. The exponential term and its conjugate is simply $2 \cos(\cdot)$ of the argument, except for the $j=1/2$ case where the argument is $0$ and there is no conjugate pair. To evaluate the argument of the cosines, we define the matrix $W_j$ such that $W_S = \textrm{diag} \left\{-1,1,0,\dots,0,1,-1 \right\}$, $W_{S-1} = \textrm{diag} \left\{ 0, -1,1,0,\dots,0,1,-1,0 \right\}$, and so forth. This yields cosine arguments of $\omega_j \tau$ where $\omega_j \equiv \textrm{Tr}(W_j \H)$, remembering that $\H$ is diagonal. The final (general) equation for the echo magnitude is then given by
\begin{equation}
    \braket{\S_x(2\tau)} = C \sum_{j=1/2}^S \lambda_j a_j \psi_j \psi_{j-1} \cos (\omega_j \tau).
\end{equation}
To help simplify, we define the prefactor variable $A^S_j = \lambda_j a^S_j \psi^S_j \psi^S_{j-1}$, and note that since $\braket{\S_x(0)} = S$ we can derive the normalization factor by setting $\tau = 0$ in the above expression, yielding $S = C \sum_{j=1/2}^S A^S_j$. This normalization is necessary as a final step, as we never normalized $\ket{\psi_0}$, and doing so from the recursive relation would be tedious.

The Hamiltonian $\H$ only enters this final expression linearly in the definitions of $\omega_j$. If we write $\H_\textrm{diag} = \H_1 + \H_2$, then the final $\omega_j = \omega_j^{(1)} + \omega_j^{(2)}$, e.g. the frequencies add linearly. Let us consider some general Hamiltonian terms then. First, we can quickly see that any identity term $\H \propto \mathbbm{1}$ must yield $\omega_j = 0$ for all $j$ (which is reassuring, as constants should not affect the dynamics of observables). Similarly, if $\H = \S_z$, its anti-symmetry yields $\omega_j = 0$. However, for $\H = \S_z^2$, the elements are $\{j^2\}$ and we have $|\omega_j| = 2(j^2 - (j-1)^2) = 4j - 2$. Now consider the quadrupolar Hamiltonian (for general $S$):
\begin{equation}
    \H_Q = \omega_Z \S_z + \frac{\omega_Q}{2} \left(3 \S_z^2 + \S^2 \right).
\end{equation}
Since $\S^2 \propto \textrm{Id}$, $\S^2$ and $\S_z$ both give no contribution to $\omega_j$, and we need only consider the $\S_z^2$ contribution:
\begin{equation}
\omega^{\H_Q}_j = 3 \omega_Q (2j - 1).
\end{equation}
We now have everything needed to derive expressions of $\braket{\S(2\tau)}$ for a thermal ensemble of general spin $S$.
We simply sum the prefactors $(C \lambda_j a_j \psi_j \psi_{j-1})$ weighted by their initial magnetization $m$.
The final results are given up to $S = 11/2$ by expressions proportional to:

\begin{equation}
\braket{\S_x(2\tau)} \propto
    \begin{cases}
        1,~&S=\tfrac{1}{2} \vspace{1.5mm} \\
        2 + 3 \cos(6 \omega_Q \tau),&S=\tfrac{3}{2} \vspace{1.5mm} \\
        9 + 16\cos(6 \omega_Q \tau) + 10 \cos(12 \omega_Q \tau),&S=\tfrac{5}{2} \vspace{1.5mm} \\
        8 + 15\cos(6 \omega_Q \tau) + 12\cos(12 \omega_Q \tau)& \\
        + 7\cos(18 \omega_Q \tau),&S=\tfrac{7}{2} \vspace{1.5mm} \\
        25 + 48\cos(6 \omega_Q \tau) + 42\cos(12 \omega_Q \tau)& \\
        + 32\cos(18 \omega_Q \tau) + 18\cos(24 \omega_Q \tau),&S=\tfrac{9}{2} \vspace{1.5mm} \\
        18 + 35\cos(6 \omega_Q \tau) + 32\cos(12 \omega_Q \tau)& \\
        + 27\cos(18 \omega_Q \tau) + 20\cos(24 \omega_Q \tau)& \\
        + 11\cos(30 \omega_Q \tau),&S=\tfrac{11}{2}.
    \end{cases}
\end{equation}

\section{Integer Spins}
\label{app:intspin}

For integer spin, the $\S_i$ matrices and initial state $\psi_0$ now have a term at their center $j=0$ that is not related to any other by symmetry.
However, the rest of the derivation is identical, and the final dot product now looks like:

\begin{equation}
\begin{split}
    \bra{2 \tau} \S_x \ket{2 \tau} = &\left[ a_S F_S^\dagger F_{S-1} \right] \\
    + &\left[ a_S F_{S-1}^\dagger F_S + a_{S-1} F_{S-1}^\dagger F_{S-2} \right] + \dots \\
    \dots + &\left[ a_{1} F_{0}^\dagger F_{1} + a_{0} F_{0}^\dagger F_{-1} \right] \\
    + &\left[ a_0 F_{-1}^\dagger F_0 + a_{-1} F_{-1}^\dagger F_{-2} \right] \\
    + &\left[ a_{-1} F_{-2}^\dagger F_{-1} + a_{-2} F_{-2}^\dagger F_{-3} \right] + \dots 
    \end{split}
\end{equation}

Unlike in the half-integer case, here every term has a hermitian conjugate. Remembering the symmetry $a_j = a_{-j+1}$ and $F_{j} = -F_{-j}^\dagger$, we can simplify as:

\begin{equation}
    \begin{split}
    \braket{\S_x (2\tau)} = & 2\sum_{j = 1}^S a_j \psi_j \psi_{j-1} \\
    \times &\left[ e^{-i(h_j + h_{-j} - h_{j-1} - h_{-j+1}) \tau} + h.c. \right] \\
    = & C \sum_{j=1}^S \tilde{A}^S_j \cos(\omega_j \tau).
    \end{split}
\end{equation}
where $\tilde{A}^S_j = a_j \psi_j \psi_{j-1}$, i.e. we no longer need the special function $\lambda_j$ to single out the unpaired term. The generating matrices $W_j$ are mostly the same (but with an extra $0$ at the center), and a special form shows up for $j=1$: $W_1 = \textrm{diag} \{ \dots, 0, -1,2,-1, 0, \dots \}$. This still leads to the same conditions: $\omega_j = 0$ for any Hamiltonian proportional to $\textrm{Id}$ or $\S_z$, and $\omega_j = 4j - 2$ for Hamiltonians proportional to $\S_z^2$. However, as $j$ is now an integer, $\omega_j \neq 0$ for any $j$. So no constant term appears in our expressions for $\braket{\S_x(2 \tau)}$.

The expressions for $S$ up to $5$ are proportional to:
\begin{equation}
\braket{\S_x(2\tau)} \propto
    \begin{cases}
        \cos(3 \omega_Q \tau),&S=1 \vspace{1.5mm} \\
        3\cos(3 \omega_Q \tau) + 2\cos(9 \omega_Q \tau),&S=2 \vspace{1.5mm} \\
        6\cos(3 \omega_Q \tau) + 5\cos(9 \omega_Q \tau)& \\
        + 3\cos(15 \omega_Q \tau),&S=3 \vspace{1.5mm} \\
        ~10\cos(3 \omega_Q \tau) + 9\cos(9 \omega_Q \tau)& \\
        + 7\cos(15 \omega_Q \tau) + 4\cos(21 \omega_Q \tau),&S=4 \vspace{1.5mm} \\
        15\cos(3 \omega_Q \tau) + 14\cos(9 \omega_Q \tau) \\
        + 12\cos(15 \omega_Q \tau) + 9\cos(21 \omega_Q \tau) \\
        + 5\cos(27 \omega_Q \tau),&S=5.
    \end{cases}
\end{equation}

%\pagebreak

\bibliography{refs}

\begin{thebibliography}{35}%
\makeatletter
\providecommand \@ifxundefined [1]{%
 \@ifx{#1\undefined}
}%
\providecommand \@ifnum [1]{%
 \ifnum #1\expandafter \@firstoftwo
 \else \expandafter \@secondoftwo
 \fi
}%
\providecommand \@ifx [1]{%
 \ifx #1\expandafter \@firstoftwo
 \else \expandafter \@secondoftwo
 \fi
}%
\providecommand \natexlab [1]{#1}%
\providecommand \enquote  [1]{``#1''}%
\providecommand \bibnamefont  [1]{#1}%
\providecommand \bibfnamefont [1]{#1}%
\providecommand \citenamefont [1]{#1}%
\providecommand \href@noop [0]{\@secondoftwo}%
\providecommand \href [0]{\begingroup \@sanitize@url \@href}%
\providecommand \@href[1]{\@@startlink{#1}\@@href}%
\providecommand \@@href[1]{\endgroup#1\@@endlink}%
\providecommand \@sanitize@url [0]{\catcode `\\12\catcode `\$12\catcode
  `\&12\catcode `\#12\catcode `\^12\catcode `\_12\catcode `\%12\relax}%
\providecommand \@@startlink[1]{}%
\providecommand \@@endlink[0]{}%
\providecommand \url  [0]{\begingroup\@sanitize@url \@url }%
\providecommand \@url [1]{\endgroup\@href {#1}{\urlprefix }}%
\providecommand \urlprefix  [0]{URL }%
\providecommand \Eprint [0]{\href }%
\providecommand \doibase [0]{https://doi.org/}%
\providecommand \selectlanguage [0]{\@gobble}%
\providecommand \bibinfo  [0]{\@secondoftwo}%
\providecommand \bibfield  [0]{\@secondoftwo}%
\providecommand \translation [1]{[#1]}%
\providecommand \BibitemOpen [0]{}%
\providecommand \bibitemStop [0]{}%
\providecommand \bibitemNoStop [0]{.\EOS\space}%
\providecommand \EOS [0]{\spacefactor3000\relax}%
\providecommand \BibitemShut  [1]{\csname bibitem#1\endcsname}%
\let\auto@bib@innerbib\@empty
%</preamble>
\bibitem [{\citenamefont {Landau}\ and\ \citenamefont
  {Lifshitz}(1980)}]{LandauBook}%
  \BibitemOpen
  \bibfield  {author} {\bibinfo {author} {\bibfnamefont {L.}~\bibnamefont
  {Landau}}\ and\ \bibinfo {author} {\bibfnamefont {E.}~\bibnamefont
  {Lifshitz}},\ }\href@noop {} {\emph {\bibinfo {title} {Statistical Physics:
  Volume 5}}}\ (\bibinfo  {publisher} {Elsevier Science},\ \bibinfo {address}
  {Boston, MA},\ \bibinfo {year} {1980})\BibitemShut {NoStop}%
\bibitem [{\citenamefont {Lipa}\ \emph {et~al.}(1996)\citenamefont {Lipa},
  \citenamefont {Swanson}, \citenamefont {Nissen}, \citenamefont {Chui},\ and\
  \citenamefont {Israelsson}}]{Lipa1996}%
  \BibitemOpen
  \bibfield  {author} {\bibinfo {author} {\bibfnamefont {J.~A.}\ \bibnamefont
  {Lipa}}, \bibinfo {author} {\bibfnamefont {D.~R.}\ \bibnamefont {Swanson}},
  \bibinfo {author} {\bibfnamefont {J.~A.}\ \bibnamefont {Nissen}}, \bibinfo
  {author} {\bibfnamefont {T.~C.~P.}\ \bibnamefont {Chui}},\ and\ \bibinfo
  {author} {\bibfnamefont {U.~E.}\ \bibnamefont {Israelsson}},\ }\bibfield
  {title} {\bibinfo {title} {Heat capacity and thermal relaxation of bulk
  helium very near the lambda point},\ }\href
  {https://doi.org/10.1103/PhysRevLett.76.944} {\bibfield  {journal} {\bibinfo
  {journal} {Phys. Rev. Lett.}\ }\textbf {\bibinfo {volume} {76}},\ \bibinfo
  {pages} {944} (\bibinfo {year} {1996})}\BibitemShut {NoStop}%
\bibitem [{\citenamefont {{Villain, J.}}\ \emph {et~al.}(1980)\citenamefont
  {{Villain, J.}}, \citenamefont {{Bidaux, R.}}, \citenamefont {{Carton,
  J.-P.}},\ and\ \citenamefont {{Conte, R.}}}]{Villain1980}%
  \BibitemOpen
  \bibfield  {author} {\bibinfo {author} {\bibnamefont {{Villain, J.}}},
  \bibinfo {author} {\bibnamefont {{Bidaux, R.}}}, \bibinfo {author}
  {\bibnamefont {{Carton, J.-P.}}},\ and\ \bibinfo {author} {\bibnamefont
  {{Conte, R.}}},\ }\bibfield  {title} {\bibinfo {title} {Order as an effect of
  disorder},\ }\href {https://doi.org/10.1051/jphys:0198000410110126300}
  {\bibfield  {journal} {\bibinfo  {journal} {J. Phys. France}\ }\textbf
  {\bibinfo {volume} {41}},\ \bibinfo {pages} {1263} (\bibinfo {year}
  {1980})}\BibitemShut {NoStop}%
\bibitem [{\citenamefont {Savary}\ \emph {et~al.}(2012)\citenamefont {Savary},
  \citenamefont {Ross}, \citenamefont {Gaulin}, \citenamefont {Ruff},\ and\
  \citenamefont {Balents}}]{Savary2012}%
  \BibitemOpen
  \bibfield  {author} {\bibinfo {author} {\bibfnamefont {L.}~\bibnamefont
  {Savary}}, \bibinfo {author} {\bibfnamefont {K.~A.}\ \bibnamefont {Ross}},
  \bibinfo {author} {\bibfnamefont {B.~D.}\ \bibnamefont {Gaulin}}, \bibinfo
  {author} {\bibfnamefont {J.~P.~C.}\ \bibnamefont {Ruff}},\ and\ \bibinfo
  {author} {\bibfnamefont {L.}~\bibnamefont {Balents}},\ }\bibfield  {title}
  {\bibinfo {title} {{Order by Quantum Disorder in Er$_2$Ti$_2$O$_7$}},\ }\href
  {https://doi.org/10.1103/PhysRevLett.109.167201} {\bibfield  {journal}
  {\bibinfo  {journal} {Phys. Rev. Lett.}\ }\textbf {\bibinfo {volume} {109}},\
  \bibinfo {pages} {167201} (\bibinfo {year} {2012})}\BibitemShut {NoStop}%
\bibitem [{\citenamefont {Sivardiere}(1973)}]{Sivardiere1973}%
  \BibitemOpen
  \bibfield  {author} {\bibinfo {author} {\bibfnamefont {J.}~\bibnamefont
  {Sivardiere}},\ }\bibfield  {title} {\bibinfo {title} {Multipolar phase
  transitions in magnetic crystals},\ }\href
  {https://doi.org/https://doi.org/10.1016/0022-3697(73)90086-3} {\bibfield
  {journal} {\bibinfo  {journal} {Journal of Physics and Chemistry of Solids}\
  }\textbf {\bibinfo {volume} {34}},\ \bibinfo {pages} {267} (\bibinfo {year}
  {1973})}\BibitemShut {NoStop}%
\bibitem [{\citenamefont {Maki}\ and\ \citenamefont
  {Nos{\'e}}(1979)}]{Maki1979}%
  \BibitemOpen
  \bibfield  {author} {\bibinfo {author} {\bibfnamefont {K.}~\bibnamefont
  {Maki}}\ and\ \bibinfo {author} {\bibfnamefont {S.}~\bibnamefont
  {Nos{\'e}}},\ }\bibfield  {title} {\bibinfo {title} {Orientational order and
  phase transitions in a two‐dimensional triangular octopolar array},\ }\href
  {https://doi.org/10.1063/1.438440} {\bibfield  {journal} {\bibinfo  {journal}
  {The Journal of Chemical Physics}\ }\textbf {\bibinfo {volume} {71}},\
  \bibinfo {pages} {1392} (\bibinfo {year} {1979})}\BibitemShut {NoStop}%
\bibitem [{\citenamefont {B{\"o}hmer}\ and\ \citenamefont
  {Loidl}(1990)}]{Bohmer1990}%
  \BibitemOpen
  \bibfield  {author} {\bibinfo {author} {\bibfnamefont {R.}~\bibnamefont
  {B{\"o}hmer}}\ and\ \bibinfo {author} {\bibfnamefont {A.}~\bibnamefont
  {Loidl}},\ }\bibfield  {title} {\bibinfo {title} {{Reorientations and phase
  transitions in (Kr)$_{1-x}$(CH$_{4-n}$D$_n$)$_x$}},\ }\href
  {https://doi.org/10.1007/BF01390661} {\bibfield  {journal} {\bibinfo
  {journal} {Zeitschrift f{\"u}r Physik B Condensed Matter}\ }\textbf {\bibinfo
  {volume} {80}},\ \bibinfo {pages} {139} (\bibinfo {year} {1990})}\BibitemShut
  {NoStop}%
\bibitem [{\citenamefont {Kiss}\ and\ \citenamefont
  {Fazekas}(2005)}]{Kiss2005}%
  \BibitemOpen
  \bibfield  {author} {\bibinfo {author} {\bibfnamefont {A.}~\bibnamefont
  {Kiss}}\ and\ \bibinfo {author} {\bibfnamefont {P.}~\bibnamefont {Fazekas}},\
  }\bibfield  {title} {\bibinfo {title} {{Group theory and octupolar order in
  $\mathrm{U}{\mathrm{Ru}}_{2}{\mathrm{Si}}_{2}$}},\ }\href
  {https://doi.org/10.1103/PhysRevB.71.054415} {\bibfield  {journal} {\bibinfo
  {journal} {Phys. Rev. B}\ }\textbf {\bibinfo {volume} {71}},\ \bibinfo
  {pages} {054415} (\bibinfo {year} {2005})}\BibitemShut {NoStop}%
\bibitem [{\citenamefont {T\'oth}\ and\ \citenamefont
  {Kotliar}(2011)}]{Toth2011}%
  \BibitemOpen
  \bibfield  {author} {\bibinfo {author} {\bibfnamefont {A.~I.}\ \bibnamefont
  {T\'oth}}\ and\ \bibinfo {author} {\bibfnamefont {G.}~\bibnamefont
  {Kotliar}},\ }\bibfield  {title} {\bibinfo {title} {{Hexadecapolar Kondo
  Effect in $\mathrm{U}{\mathrm{Ru}}_{2}\mathrm{S}{\mathrm{i}}_{2}$?}},\ }\href
  {https://doi.org/10.1103/PhysRevLett.107.266405} {\bibfield  {journal}
  {\bibinfo  {journal} {Phys. Rev. Lett.}\ }\textbf {\bibinfo {volume} {107}},\
  \bibinfo {pages} {266405} (\bibinfo {year} {2011})}\BibitemShut {NoStop}%
\bibitem [{\citenamefont {Senyuk}\ \emph {et~al.}(2016)\citenamefont {Senyuk},
  \citenamefont {Puls}, \citenamefont {Tovkach}, \citenamefont {Chernyshuk},\
  and\ \citenamefont {Smalyukh}}]{Senyuk2016}%
  \BibitemOpen
  \bibfield  {author} {\bibinfo {author} {\bibfnamefont {B.}~\bibnamefont
  {Senyuk}}, \bibinfo {author} {\bibfnamefont {O.}~\bibnamefont {Puls}},
  \bibinfo {author} {\bibfnamefont {O.~M.}\ \bibnamefont {Tovkach}}, \bibinfo
  {author} {\bibfnamefont {S.~B.}\ \bibnamefont {Chernyshuk}},\ and\ \bibinfo
  {author} {\bibfnamefont {I.~I.}\ \bibnamefont {Smalyukh}},\ }\bibfield
  {title} {\bibinfo {title} {Hexadecapolar colloids},\ }\href
  {https://doi.org/10.1038/ncomms10659} {\bibfield  {journal} {\bibinfo
  {journal} {Nature Communications}\ }\textbf {\bibinfo {volume} {7}},\
  \bibinfo {pages} {10659} (\bibinfo {year} {2016})}\BibitemShut {NoStop}%
\bibitem [{\citenamefont {Caciuffo}\ \emph {et~al.}(2003)\citenamefont
  {Caciuffo}, \citenamefont {o}, \citenamefont {Detlefs}, \citenamefont
  {Longfield}, \citenamefont {Santini}, \citenamefont {Bernhoeft},
  \citenamefont {Rebizant},\ and\ \citenamefont {Lander}}]{Caciuffo2003}%
  \BibitemOpen
  \bibfield  {author} {\bibinfo {author} {\bibfnamefont {R.}~\bibnamefont
  {Caciuffo}}, \bibinfo {author} {\bibfnamefont {J.~A.~P.}\ \bibnamefont {o}},
  \bibinfo {author} {\bibfnamefont {C.}~\bibnamefont {Detlefs}}, \bibinfo
  {author} {\bibfnamefont {M.~J.}\ \bibnamefont {Longfield}}, \bibinfo {author}
  {\bibfnamefont {P.}~\bibnamefont {Santini}}, \bibinfo {author} {\bibfnamefont
  {N.}~\bibnamefont {Bernhoeft}}, \bibinfo {author} {\bibfnamefont
  {J.}~\bibnamefont {Rebizant}},\ and\ \bibinfo {author} {\bibfnamefont
  {G.~H.}\ \bibnamefont {Lander}},\ }\bibfield  {title} {\bibinfo {title}
  {{Multipolar ordering in {NpO}$_2$ below 25 K}},\ }\href
  {https://doi.org/10.1088/0953-8984/15/28/370} {\bibfield  {journal} {\bibinfo
   {journal} {Journal of Physics: Condensed Matter}\ }\textbf {\bibinfo
  {volume} {15}},\ \bibinfo {pages} {S2287} (\bibinfo {year}
  {2003})}\BibitemShut {NoStop}%
\bibitem [{\citenamefont {Onimaru}\ \emph {et~al.}(2005)\citenamefont
  {Onimaru}, \citenamefont {Sakakibara}, \citenamefont {Aso}, \citenamefont
  {Yoshizawa}, \citenamefont {Suzuki},\ and\ \citenamefont
  {Takeuchi}}]{Onimaru2005}%
  \BibitemOpen
  \bibfield  {author} {\bibinfo {author} {\bibfnamefont {T.}~\bibnamefont
  {Onimaru}}, \bibinfo {author} {\bibfnamefont {T.}~\bibnamefont {Sakakibara}},
  \bibinfo {author} {\bibfnamefont {N.}~\bibnamefont {Aso}}, \bibinfo {author}
  {\bibfnamefont {H.}~\bibnamefont {Yoshizawa}}, \bibinfo {author}
  {\bibfnamefont {H.~S.}\ \bibnamefont {Suzuki}},\ and\ \bibinfo {author}
  {\bibfnamefont {T.}~\bibnamefont {Takeuchi}},\ }\bibfield  {title} {\bibinfo
  {title} {{Observation of Modulated Quadrupolar Structures in
  ${\mathrm{PrPb}}_{3}$}},\ }\href
  {https://doi.org/10.1103/PhysRevLett.94.197201} {\bibfield  {journal}
  {\bibinfo  {journal} {Phys. Rev. Lett.}\ }\textbf {\bibinfo {volume} {94}},\
  \bibinfo {pages} {197201} (\bibinfo {year} {2005})}\BibitemShut {NoStop}%
\bibitem [{\citenamefont {Bombardi}\ \emph {et~al.}(2008)\citenamefont
  {Bombardi}, \citenamefont {Mazzoli}, \citenamefont {Agrestini},\ and\
  \citenamefont {Lees}}]{Bombardi2008}%
  \BibitemOpen
  \bibfield  {author} {\bibinfo {author} {\bibfnamefont {A.}~\bibnamefont
  {Bombardi}}, \bibinfo {author} {\bibfnamefont {C.}~\bibnamefont {Mazzoli}},
  \bibinfo {author} {\bibfnamefont {S.}~\bibnamefont {Agrestini}},\ and\
  \bibinfo {author} {\bibfnamefont {M.~R.}\ \bibnamefont {Lees}},\ }\bibfield
  {title} {\bibinfo {title} {{Resonant x-ray scattering investigation of the
  multipolar ordering in ${\text{Ca}}_{3}{\text{Co}}_{2}{\text{O}}_{6}$}},\
  }\href {https://doi.org/10.1103/PhysRevB.78.100406} {\bibfield  {journal}
  {\bibinfo  {journal} {Phys. Rev. B}\ }\textbf {\bibinfo {volume} {78}},\
  \bibinfo {pages} {100406} (\bibinfo {year} {2008})}\BibitemShut {NoStop}%
\bibitem [{\citenamefont {Shen}\ \emph {et~al.}(2019)\citenamefont {Shen},
  \citenamefont {Liu}, \citenamefont {Qin}, \citenamefont {Shen}, \citenamefont
  {Li}, \citenamefont {Bewley}, \citenamefont {Schneidewind}, \citenamefont
  {Chen},\ and\ \citenamefont {Zhao}}]{Shen2019}%
  \BibitemOpen
  \bibfield  {author} {\bibinfo {author} {\bibfnamefont {Y.}~\bibnamefont
  {Shen}}, \bibinfo {author} {\bibfnamefont {C.}~\bibnamefont {Liu}}, \bibinfo
  {author} {\bibfnamefont {Y.}~\bibnamefont {Qin}}, \bibinfo {author}
  {\bibfnamefont {S.}~\bibnamefont {Shen}}, \bibinfo {author} {\bibfnamefont
  {Y.-D.}\ \bibnamefont {Li}}, \bibinfo {author} {\bibfnamefont
  {R.}~\bibnamefont {Bewley}}, \bibinfo {author} {\bibfnamefont
  {A.}~\bibnamefont {Schneidewind}}, \bibinfo {author} {\bibfnamefont
  {G.}~\bibnamefont {Chen}},\ and\ \bibinfo {author} {\bibfnamefont
  {J.}~\bibnamefont {Zhao}},\ }\bibfield  {title} {\bibinfo {title}
  {{Intertwined dipolar and multipolar order in the triangular-lattice magnet
  TmMgGaO$_4$}},\ }\href {https://doi.org/10.1038/s41467-019-12410-3}
  {\bibfield  {journal} {\bibinfo  {journal} {Nature Communications}\ }\textbf
  {\bibinfo {volume} {10}},\ \bibinfo {pages} {4530} (\bibinfo {year}
  {2019})}\BibitemShut {NoStop}%
\bibitem [{\citenamefont {Maharaj}\ \emph {et~al.}(2020)\citenamefont
  {Maharaj}, \citenamefont {Sala}, \citenamefont {Stone}, \citenamefont
  {Kermarrec}, \citenamefont {Ritter}, \citenamefont {Fauth}, \citenamefont
  {Marjerrison}, \citenamefont {Greedan}, \citenamefont {Paramekanti},\ and\
  \citenamefont {Gaulin}}]{Maharaj2020}%
  \BibitemOpen
  \bibfield  {author} {\bibinfo {author} {\bibfnamefont {D.~D.}\ \bibnamefont
  {Maharaj}}, \bibinfo {author} {\bibfnamefont {G.}~\bibnamefont {Sala}},
  \bibinfo {author} {\bibfnamefont {M.~B.}\ \bibnamefont {Stone}}, \bibinfo
  {author} {\bibfnamefont {E.}~\bibnamefont {Kermarrec}}, \bibinfo {author}
  {\bibfnamefont {C.}~\bibnamefont {Ritter}}, \bibinfo {author} {\bibfnamefont
  {F.}~\bibnamefont {Fauth}}, \bibinfo {author} {\bibfnamefont {C.~A.}\
  \bibnamefont {Marjerrison}}, \bibinfo {author} {\bibfnamefont {J.~E.}\
  \bibnamefont {Greedan}}, \bibinfo {author} {\bibfnamefont {A.}~\bibnamefont
  {Paramekanti}},\ and\ \bibinfo {author} {\bibfnamefont {B.~D.}\ \bibnamefont
  {Gaulin}},\ }\bibfield  {title} {\bibinfo {title} {{Octupolar versus N\'eel
  Order in Cubic $5{d}^{2}$ Double Perovskites}},\ }\href
  {https://doi.org/10.1103/PhysRevLett.124.087206} {\bibfield  {journal}
  {\bibinfo  {journal} {Phys. Rev. Lett.}\ }\textbf {\bibinfo {volume} {124}},\
  \bibinfo {pages} {087206} (\bibinfo {year} {2020})}\BibitemShut {NoStop}%
\bibitem [{\citenamefont {Proffen}\ \emph {et~al.}(2003)\citenamefont
  {Proffen}, \citenamefont {Billinge}, \citenamefont {Egami},\ and\
  \citenamefont {Louca}}]{Proffen2003}%
  \BibitemOpen
  \bibfield  {author} {\bibinfo {author} {\bibfnamefont {T.}~\bibnamefont
  {Proffen}}, \bibinfo {author} {\bibfnamefont {S.~J.~L.}\ \bibnamefont
  {Billinge}}, \bibinfo {author} {\bibfnamefont {T.}~\bibnamefont {Egami}},\
  and\ \bibinfo {author} {\bibfnamefont {D.}~\bibnamefont {Louca}},\ }\bibfield
   {title} {\bibinfo {title} {Structural analysis of complex materials using
  the atomic pair distribution function --- a practical guide},\ }\href
  {https://doi.org/doi:10.1524/zkri.218.2.132.20664} {\bibfield  {journal}
  {\bibinfo  {journal} {Zeitschrift f{\"u}r Kristallographie - Crystalline
  Materials}\ }\textbf {\bibinfo {volume} {218}},\ \bibinfo {pages} {132}
  (\bibinfo {year} {2003})}\BibitemShut {NoStop}%
\bibitem [{\citenamefont {Fornasini}\ and\ \citenamefont
  {Grisenti}(2015)}]{Fornasini2015}%
  \BibitemOpen
  \bibfield  {author} {\bibinfo {author} {\bibfnamefont {P.}~\bibnamefont
  {Fornasini}}\ and\ \bibinfo {author} {\bibfnamefont {R.}~\bibnamefont
  {Grisenti}},\ }\bibfield  {title} {\bibinfo {title} {{On EXAFS Debye-Waller
  factor and recent advances}},\ }\href
  {https://doi.org/10.1107/S1600577515010759} {\bibfield  {journal} {\bibinfo
  {journal} {Journal of Synchrotron Radiation}\ }\textbf {\bibinfo {volume}
  {22}},\ \bibinfo {pages} {1242} (\bibinfo {year} {2015})}\BibitemShut
  {NoStop}%
\bibitem [{\citenamefont {Welberry}\ and\ \citenamefont
  {Weber}(2016)}]{Welberry2016}%
  \BibitemOpen
  \bibfield  {author} {\bibinfo {author} {\bibfnamefont {T.}~\bibnamefont
  {Welberry}}\ and\ \bibinfo {author} {\bibfnamefont {T.}~\bibnamefont
  {Weber}},\ }\bibfield  {title} {\bibinfo {title} {One hundred years of
  diffuse scattering},\ }\href {https://doi.org/10.1080/0889311X.2015.1046853}
  {\bibfield  {journal} {\bibinfo  {journal} {Crystallography Reviews}\
  }\textbf {\bibinfo {volume} {22}},\ \bibinfo {pages} {2} (\bibinfo {year}
  {2016})}\BibitemShut {NoStop}%
\bibitem [{\citenamefont {Pourovskii}\ \emph {et~al.}(2021)\citenamefont
  {Pourovskii}, \citenamefont {Mosca},\ and\ \citenamefont
  {Franchini}}]{Pourovskii2021}%
  \BibitemOpen
  \bibfield  {author} {\bibinfo {author} {\bibfnamefont {L.~V.}\ \bibnamefont
  {Pourovskii}}, \bibinfo {author} {\bibfnamefont {D.~F.}\ \bibnamefont
  {Mosca}},\ and\ \bibinfo {author} {\bibfnamefont {C.}~\bibnamefont
  {Franchini}},\ }\bibfield  {title} {\bibinfo {title} {{Ferro-octupolar Order
  and Low-Energy Excitations in ${\mathrm{d}}^{2}$ Double Perovskites of
  Osmium}},\ }\href {https://doi.org/10.1103/PhysRevLett.127.237201} {\bibfield
   {journal} {\bibinfo  {journal} {Phys. Rev. Lett.}\ }\textbf {\bibinfo
  {volume} {127}},\ \bibinfo {pages} {237201} (\bibinfo {year}
  {2021})}\BibitemShut {NoStop}%
\bibitem [{\citenamefont {Zaliznyak}\ and\ \citenamefont
  {Tranquada}(2015)}]{Zaliznyak2015}%
  \BibitemOpen
  \bibfield  {author} {\bibinfo {author} {\bibfnamefont {I.~A.}\ \bibnamefont
  {Zaliznyak}}\ and\ \bibinfo {author} {\bibfnamefont {J.~M.}\ \bibnamefont
  {Tranquada}},\ }\bibinfo {title} {Neutron scattering and its application to
  strongly correlated systems},\ in\ \href
  {https://doi.org/10.1007/978-3-662-44133-6_7} {\emph {\bibinfo {booktitle}
  {Strongly Correlated Systems: Experimental Techniques}}},\ \bibinfo {editor}
  {edited by\ \bibinfo {editor} {\bibfnamefont {A.}~\bibnamefont {Avella}}\
  and\ \bibinfo {editor} {\bibfnamefont {F.}~\bibnamefont {Mancini}}}\
  (\bibinfo  {publisher} {Springer Berlin Heidelberg},\ \bibinfo {address}
  {Berlin, Heidelberg},\ \bibinfo {year} {2015})\ pp.\ \bibinfo {pages}
  {205--235}\BibitemShut {NoStop}%
\bibitem [{\citenamefont {Brown}\ and\ \citenamefont
  {Wimperis}(1997)}]{Brown1997}%
  \BibitemOpen
  \bibfield  {author} {\bibinfo {author} {\bibfnamefont {S.~P.}\ \bibnamefont
  {Brown}}\ and\ \bibinfo {author} {\bibfnamefont {S.}~\bibnamefont
  {Wimperis}},\ }\bibfield  {title} {\bibinfo {title} {{Two-Dimensional
  Multiple-Quantum MAS NMR of Quadrupolar Nuclei: A Comparison of Methods}},\
  }\href {https://doi.org/https://doi.org/10.1006/jmre.1997.1217} {\bibfield
  {journal} {\bibinfo  {journal} {Journal of Magnetic Resonance}\ }\textbf
  {\bibinfo {volume} {128}},\ \bibinfo {pages} {42} (\bibinfo {year}
  {1997})}\BibitemShut {NoStop}%
\bibitem [{\citenamefont {M{\'a}di}\ \emph {et~al.}(1998)\citenamefont
  {M{\'a}di}, \citenamefont {Br{\"u}schweiler},\ and\ \citenamefont
  {Ernst}}]{Madi1998}%
  \BibitemOpen
  \bibfield  {author} {\bibinfo {author} {\bibfnamefont {Z.~L.}\ \bibnamefont
  {M{\'a}di}}, \bibinfo {author} {\bibfnamefont {R.}~\bibnamefont
  {Br{\"u}schweiler}},\ and\ \bibinfo {author} {\bibfnamefont {R.~R.}\
  \bibnamefont {Ernst}},\ }\bibfield  {title} {\bibinfo {title} {One- and
  two-dimensional ensemble quantum computing in spin liouville space},\ }\href
  {https://doi.org/10.1063/1.477759} {\bibfield  {journal} {\bibinfo  {journal}
  {The Journal of Chemical Physics}\ }\textbf {\bibinfo {volume} {109}},\
  \bibinfo {pages} {10603} (\bibinfo {year} {1998})}\BibitemShut {NoStop}%
\bibitem [{\citenamefont {Smallwood}\ and\ \citenamefont
  {Cundiff}(2018)}]{Smallwood2018}%
  \BibitemOpen
  \bibfield  {author} {\bibinfo {author} {\bibfnamefont {C.~L.}\ \bibnamefont
  {Smallwood}}\ and\ \bibinfo {author} {\bibfnamefont {S.~T.}\ \bibnamefont
  {Cundiff}},\ }\bibfield  {title} {\bibinfo {title} {Multidimensional coherent
  spectroscopy of semiconductors},\ }\href
  {https://doi.org/https://doi.org/10.1002/lpor.201800171} {\bibfield
  {journal} {\bibinfo  {journal} {Laser \& Photonics Reviews}\ }\textbf
  {\bibinfo {volume} {12}},\ \bibinfo {pages} {1800171} (\bibinfo {year}
  {2018})}\BibitemShut {NoStop}%
\bibitem [{\citenamefont {M{\"u}ller}\ \emph {et~al.}(1975)\citenamefont
  {M{\"u}ller}, \citenamefont {Kumar},\ and\ \citenamefont
  {Ernst}}]{Muller1975}%
  \BibitemOpen
  \bibfield  {author} {\bibinfo {author} {\bibfnamefont {L.}~\bibnamefont
  {M{\"u}ller}}, \bibinfo {author} {\bibfnamefont {A.}~\bibnamefont {Kumar}},\
  and\ \bibinfo {author} {\bibfnamefont {R.~R.}\ \bibnamefont {Ernst}},\
  }\bibfield  {title} {\bibinfo {title} {{Two‐dimensional carbon‐13 NMR
  spectroscopy}},\ }\href {https://doi.org/10.1063/1.431284} {\bibfield
  {journal} {\bibinfo  {journal} {The Journal of Chemical Physics}\ }\textbf
  {\bibinfo {volume} {63}},\ \bibinfo {pages} {5490} (\bibinfo {year}
  {1975})}\BibitemShut {NoStop}%
\bibitem [{\citenamefont {Lemmer}\ \emph {et~al.}(2015)\citenamefont {Lemmer},
  \citenamefont {Cormick}, \citenamefont {Schmiegelow}, \citenamefont
  {Schmidt-Kaler},\ and\ \citenamefont {Plenio}}]{Lemmer2015}%
  \BibitemOpen
  \bibfield  {author} {\bibinfo {author} {\bibfnamefont {A.}~\bibnamefont
  {Lemmer}}, \bibinfo {author} {\bibfnamefont {C.}~\bibnamefont {Cormick}},
  \bibinfo {author} {\bibfnamefont {C.~T.}\ \bibnamefont {Schmiegelow}},
  \bibinfo {author} {\bibfnamefont {F.}~\bibnamefont {Schmidt-Kaler}},\ and\
  \bibinfo {author} {\bibfnamefont {M.~B.}\ \bibnamefont {Plenio}},\ }\bibfield
   {title} {\bibinfo {title} {{Two-Dimensional Spectroscopy for the Study of
  Ion Coulomb Crystals}},\ }\href
  {https://doi.org/10.1103/PhysRevLett.114.073001} {\bibfield  {journal}
  {\bibinfo  {journal} {Phys. Rev. Lett.}\ }\textbf {\bibinfo {volume} {114}},\
  \bibinfo {pages} {073001} (\bibinfo {year} {2015})}\BibitemShut {NoStop}%
\bibitem [{\citenamefont {Sza{\'{n}}kowski}\ \emph {et~al.}(2017)\citenamefont
  {Sza{\'{n}}kowski}, \citenamefont {Ramon}, \citenamefont {Krzywda},
  \citenamefont {Kwiatkowski},\ and\ \citenamefont
  {Cywi{\'{n}}ski}}]{Szakowski2017}%
  \BibitemOpen
  \bibfield  {author} {\bibinfo {author} {\bibfnamefont {P.}~\bibnamefont
  {Sza{\'{n}}kowski}}, \bibinfo {author} {\bibfnamefont {G.}~\bibnamefont
  {Ramon}}, \bibinfo {author} {\bibfnamefont {J.}~\bibnamefont {Krzywda}},
  \bibinfo {author} {\bibfnamefont {D.}~\bibnamefont {Kwiatkowski}},\ and\
  \bibinfo {author} {\bibfnamefont {{\L}.}~\bibnamefont {Cywi{\'{n}}ski}},\
  }\bibfield  {title} {\bibinfo {title} {Environmental noise spectroscopy with
  qubits subjected to dynamical decoupling},\ }\href
  {https://doi.org/10.1088/1361-648x/aa7648} {\bibfield  {journal} {\bibinfo
  {journal} {Journal of Physics: Condensed Matter}\ }\textbf {\bibinfo {volume}
  {29}},\ \bibinfo {pages} {333001} (\bibinfo {year} {2017})}\BibitemShut
  {NoStop}%
\bibitem [{\citenamefont {Sung}\ \emph {et~al.}(2021)\citenamefont {Sung},
  \citenamefont {Veps{\"a}l{\"a}inen}, \citenamefont {Braum{\"u}ller},
  \citenamefont {Yan}, \citenamefont {Wang}, \citenamefont {Kjaergaard},
  \citenamefont {Winik}, \citenamefont {Krantz}, \citenamefont {Bengtsson},
  \citenamefont {Melville}, \citenamefont {Niedzielski}, \citenamefont
  {Schwartz}, \citenamefont {Kim}, \citenamefont {Yoder}, \citenamefont
  {Orlando}, \citenamefont {Gustavsson},\ and\ \citenamefont
  {Oliver}}]{Sung2021}%
  \BibitemOpen
  \bibfield  {author} {\bibinfo {author} {\bibfnamefont {Y.}~\bibnamefont
  {Sung}}, \bibinfo {author} {\bibfnamefont {A.}~\bibnamefont
  {Veps{\"a}l{\"a}inen}}, \bibinfo {author} {\bibfnamefont {J.}~\bibnamefont
  {Braum{\"u}ller}}, \bibinfo {author} {\bibfnamefont {F.}~\bibnamefont {Yan}},
  \bibinfo {author} {\bibfnamefont {J.~I.-J.}\ \bibnamefont {Wang}}, \bibinfo
  {author} {\bibfnamefont {M.}~\bibnamefont {Kjaergaard}}, \bibinfo {author}
  {\bibfnamefont {R.}~\bibnamefont {Winik}}, \bibinfo {author} {\bibfnamefont
  {P.}~\bibnamefont {Krantz}}, \bibinfo {author} {\bibfnamefont
  {A.}~\bibnamefont {Bengtsson}}, \bibinfo {author} {\bibfnamefont {A.~J.}\
  \bibnamefont {Melville}}, \bibinfo {author} {\bibfnamefont {B.~M.}\
  \bibnamefont {Niedzielski}}, \bibinfo {author} {\bibfnamefont {M.~E.}\
  \bibnamefont {Schwartz}}, \bibinfo {author} {\bibfnamefont {D.~K.}\
  \bibnamefont {Kim}}, \bibinfo {author} {\bibfnamefont {J.~L.}\ \bibnamefont
  {Yoder}}, \bibinfo {author} {\bibfnamefont {T.~P.}\ \bibnamefont {Orlando}},
  \bibinfo {author} {\bibfnamefont {S.}~\bibnamefont {Gustavsson}},\ and\
  \bibinfo {author} {\bibfnamefont {W.~D.}\ \bibnamefont {Oliver}},\ }\bibfield
   {title} {\bibinfo {title} {Multi-level quantum noise spectroscopy},\ }\href
  {https://doi.org/10.1038/s41467-021-21098-3} {\bibfield  {journal} {\bibinfo
  {journal} {Nature Communications}\ }\textbf {\bibinfo {volume} {12}},\
  \bibinfo {pages} {967} (\bibinfo {year} {2021})}\BibitemShut {NoStop}%
\bibitem [{\citenamefont {Cho}\ \emph {et~al.}(2005)\citenamefont {Cho},
  \citenamefont {Ladd}, \citenamefont {Baugh}, \citenamefont {Cory},\ and\
  \citenamefont {Ramanathan}}]{Cho2005}%
  \BibitemOpen
  \bibfield  {author} {\bibinfo {author} {\bibfnamefont {H.}~\bibnamefont
  {Cho}}, \bibinfo {author} {\bibfnamefont {T.~D.}\ \bibnamefont {Ladd}},
  \bibinfo {author} {\bibfnamefont {J.}~\bibnamefont {Baugh}}, \bibinfo
  {author} {\bibfnamefont {D.~G.}\ \bibnamefont {Cory}},\ and\ \bibinfo
  {author} {\bibfnamefont {C.}~\bibnamefont {Ramanathan}},\ }\bibfield  {title}
  {\bibinfo {title} {Multispin dynamics of the solid-state nmr free induction
  decay},\ }\href {https://doi.org/10.1103/PhysRevB.72.054427} {\bibfield
  {journal} {\bibinfo  {journal} {Phys. Rev. B}\ }\textbf {\bibinfo {volume}
  {72}},\ \bibinfo {pages} {054427} (\bibinfo {year} {2005})}\BibitemShut
  {NoStop}%
\bibitem [{\citenamefont {Cho}\ \emph {et~al.}(2006)\citenamefont {Cho},
  \citenamefont {Cappellaro}, \citenamefont {Cory},\ and\ \citenamefont
  {Ramanathan}}]{Cho2006}%
  \BibitemOpen
  \bibfield  {author} {\bibinfo {author} {\bibfnamefont {H.}~\bibnamefont
  {Cho}}, \bibinfo {author} {\bibfnamefont {P.}~\bibnamefont {Cappellaro}},
  \bibinfo {author} {\bibfnamefont {D.~G.}\ \bibnamefont {Cory}},\ and\
  \bibinfo {author} {\bibfnamefont {C.}~\bibnamefont {Ramanathan}},\ }\bibfield
   {title} {\bibinfo {title} {Decay of highly correlated spin states in a
  dipolar-coupled solid: Nmr study of $\mathrm{Ca}{\mathrm{f}}_{2}$},\ }\href
  {https://doi.org/10.1103/PhysRevB.74.224434} {\bibfield  {journal} {\bibinfo
  {journal} {Phys. Rev. B}\ }\textbf {\bibinfo {volume} {74}},\ \bibinfo
  {pages} {224434} (\bibinfo {year} {2006})}\BibitemShut {NoStop}%
\bibitem [{\citenamefont {Pound}(1950)}]{Pound1950}%
  \BibitemOpen
  \bibfield  {author} {\bibinfo {author} {\bibfnamefont {R.~V.}\ \bibnamefont
  {Pound}},\ }\bibfield  {title} {\bibinfo {title} {Nuclear electric quadrupole
  interactions in crystals},\ }\href {https://doi.org/10.1103/PhysRev.79.685}
  {\bibfield  {journal} {\bibinfo  {journal} {Phys. Rev.}\ }\textbf {\bibinfo
  {volume} {79}},\ \bibinfo {pages} {685} (\bibinfo {year} {1950})}\BibitemShut
  {NoStop}%
\bibitem [{\citenamefont {Antonijevic}\ and\ \citenamefont
  {Wimperis}(2005)}]{Antonijevic2005}%
  \BibitemOpen
  \bibfield  {author} {\bibinfo {author} {\bibfnamefont {S.}~\bibnamefont
  {Antonijevic}}\ and\ \bibinfo {author} {\bibfnamefont {S.}~\bibnamefont
  {Wimperis}},\ }\bibfield  {title} {\bibinfo {title} {{Separation of
  quadrupolar and chemical/paramagnetic shift interactions in two-dimensional
  $^2$H(I=1) nuclear magnetic resonance spectroscopy}},\ }\href
  {https://doi.org/10.1063/1.1807814} {\bibfield  {journal} {\bibinfo
  {journal} {The Journal of Chemical Physics}\ }\textbf {\bibinfo {volume}
  {122}},\ \bibinfo {pages} {044312} (\bibinfo {year} {2005})}\BibitemShut
  {NoStop}%
\bibitem [{\citenamefont {Abe}\ \emph {et~al.}(1966)\citenamefont {Abe},
  \citenamefont {Yasuoka},\ and\ \citenamefont {Hirai}}]{Abe1966}%
  \BibitemOpen
  \bibfield  {author} {\bibinfo {author} {\bibfnamefont {H.}~\bibnamefont
  {Abe}}, \bibinfo {author} {\bibfnamefont {H.}~\bibnamefont {Yasuoka}},\ and\
  \bibinfo {author} {\bibfnamefont {A.}~\bibnamefont {Hirai}},\ }\bibfield
  {title} {\bibinfo {title} {Spin echo modulation caused by the quadrupole
  interaction and multiple spin echoes},\ }\href
  {https://doi.org/10.1143/JPSJ.21.77} {\bibfield  {journal} {\bibinfo
  {journal} {Journal of the Physical Society of Japan}\ }\textbf {\bibinfo
  {volume} {21}},\ \bibinfo {pages} {77} (\bibinfo {year} {1966})}\BibitemShut
  {NoStop}%
\bibitem [{\citenamefont {Abe}\ \emph {et~al.}(1964)\citenamefont {Abe},
  \citenamefont {Yasuoka}, \citenamefont {Matsuura}, \citenamefont {Hirai},\
  and\ \citenamefont {Shinjo}}]{Abe1964}%
  \BibitemOpen
  \bibfield  {author} {\bibinfo {author} {\bibfnamefont {H.}~\bibnamefont
  {Abe}}, \bibinfo {author} {\bibfnamefont {H.}~\bibnamefont {Yasuoka}},
  \bibinfo {author} {\bibfnamefont {M.}~\bibnamefont {Matsuura}}, \bibinfo
  {author} {\bibfnamefont {A.}~\bibnamefont {Hirai}},\ and\ \bibinfo {author}
  {\bibfnamefont {T.}~\bibnamefont {Shinjo}},\ }\bibfield  {title} {\bibinfo
  {title} {Spin {Echo} {Modulation} {Caused} by the {Quadrupole} {Interaction}
  {Observed} on {Boron} {Nuclei} in {Ferromagnetic} {Fe}$_2${B}},\ }\href
  {https://doi.org/10.1143/JPSJ.19.1491} {\bibfield  {journal} {\bibinfo
  {journal} {Journal of the Physical Society of Japan}\ }\textbf {\bibinfo
  {volume} {19}},\ \bibinfo {pages} {1491} (\bibinfo {year}
  {1964})}\BibitemShut {NoStop}%
\bibitem [{\citenamefont {Vachon}\ \emph {et~al.}(2006)\citenamefont {Vachon},
  \citenamefont {Kundhikanjana}, \citenamefont {Straub}, \citenamefont
  {Mitrovi{\'{c}}}, \citenamefont {Reyes}, \citenamefont {Kuhns}, \citenamefont
  {Coldea},\ and\ \citenamefont {Tylczynski}}]{Vachon2006}%
  \BibitemOpen
  \bibfield  {author} {\bibinfo {author} {\bibfnamefont {M.-A.}\ \bibnamefont
  {Vachon}}, \bibinfo {author} {\bibfnamefont {W.}~\bibnamefont
  {Kundhikanjana}}, \bibinfo {author} {\bibfnamefont {A.}~\bibnamefont
  {Straub}}, \bibinfo {author} {\bibfnamefont {V.~F.}\ \bibnamefont
  {Mitrovi{\'{c}}}}, \bibinfo {author} {\bibfnamefont {A.~P.}\ \bibnamefont
  {Reyes}}, \bibinfo {author} {\bibfnamefont {P.}~\bibnamefont {Kuhns}},
  \bibinfo {author} {\bibfnamefont {R.}~\bibnamefont {Coldea}},\ and\ \bibinfo
  {author} {\bibfnamefont {Z.}~\bibnamefont {Tylczynski}},\ }\bibfield  {title}
  {\bibinfo {title} {{$^{133}$Cs {NMR} investigation of 2D frustrated
  Heisenberg antiferromagnet, Cs$_2$CuCl$_4$}},\ }\href
  {https://doi.org/10.1088/1367-2630/8/10/222} {\bibfield  {journal} {\bibinfo
  {journal} {New Journal of Physics}\ }\textbf {\bibinfo {volume} {8}},\
  \bibinfo {pages} {222} (\bibinfo {year} {2006})}\BibitemShut {NoStop}%
\bibitem [{\citenamefont {Reif}\ \emph {et~al.}(2021)\citenamefont {Reif},
  \citenamefont {Ashbrook}, \citenamefont {Emsley},\ and\ \citenamefont
  {Hong}}]{Reif2021}%
  \BibitemOpen
  \bibfield  {author} {\bibinfo {author} {\bibfnamefont {B.}~\bibnamefont
  {Reif}}, \bibinfo {author} {\bibfnamefont {S.~E.}\ \bibnamefont {Ashbrook}},
  \bibinfo {author} {\bibfnamefont {L.}~\bibnamefont {Emsley}},\ and\ \bibinfo
  {author} {\bibfnamefont {M.}~\bibnamefont {Hong}},\ }\bibfield  {title}
  {\bibinfo {title} {Solid-state nmr spectroscopy},\ }\href
  {https://doi.org/10.1038/s43586-020-00002-1} {\bibfield  {journal} {\bibinfo
  {journal} {Nature Reviews Methods Primers}\ }\textbf {\bibinfo {volume}
  {1}},\ \bibinfo {pages} {2} (\bibinfo {year} {2021})}\BibitemShut {NoStop}%
\end{thebibliography}%

\end{document}